\documentclass[%
reprint,
superscriptaddress,
 amsmath,amssymb,
 aps,
 prd,
]{revtex4-1}

\usepackage{enumitem}
\usepackage{graphicx}
\usepackage{amssymb} % Need some of symbols including <~
\usepackage{xspace} 
\usepackage{verbatim} 
\usepackage[normalem]{ulem}

\usepackage{hyperref}

\hypersetup{
    colorlinks=true,
    linkcolor=black,
    citecolor=black,
    filecolor=black,
    urlcolor=black,
}

\usepackage{color} % just used to highlight text that still needs fixed in red

\newcommand{\znu}{$0\nu\beta\beta$\xspace}
\newcommand{\tnu}{$2\nu\beta\beta$\xspace}
\newcommand{\natXe}{$\mathrm{^{nat}Xe}$\xspace}
\newcommand{\isoXe}{$\mathrm{^{136}Xe}$\xspace}
\newcommand{\enrXe}{$\mathrm{^{enr}Xe}$\xspace}
\newcommand{\meff}{$\langle m_{\beta\beta} \rangle$\xspace}

\begin{document}

\title{Kiloton-scale xenon detectors for neutrinoless double beta decay\\ and other new physics searches}

\author{A.~Avasthi}\affiliation{Department of Physics, Case Western Reserve University, Cleveland, OH 44106, USA}
\author{T.W.~Bowyer}\affiliation{Pacific Northwest National Laboratory, Richland, WA 99352, USA}
\author{C.~Bray}\affiliation{Department of Physics, Colorado School of Mines, Golden, CO 80401, USA}
\author{T.~Brunner}\affiliation{Physics Department, McGill University, Montr\'eal, Qu\'ebec H3A 2T8, Canada}\affiliation{TRIUMF, Vancouver, British Columbia V6T 2A3, Canada}
\author{N. Catarineu}\affiliation{Lawrence Livermore National Laboratory, Livermore, CA 94550, USA}
\author{E.~Church}\affiliation{Pacific Northwest National Laboratory, Richland, WA 99352, USA}
\author{R.~Guenette}\affiliation{Department of Physics, Harvard University, Cambridge, Massachusetts, USA}
\author{S.J.~Haselschwardt}\affiliation{Lawrence Berkeley National Laboratory (LBNL), Berkeley, CA 94720, USA}
\author{J.C.~Hayes}\affiliation{Pacific Northwest National Laboratory, Richland, WA 99352, USA}
\author{M.~Heffner}\email{mheffner@llnl.gov}\affiliation{Lawrence Livermore National Laboratory, Livermore, CA 94550, USA}
\author{S.A.~Hertel}\affiliation{Amherst Center for Fundamental Interactions and Physics Department, University of Massachusetts, Amherst, MA 01003, USA}
\author{P.H. Humble}\affiliation{Pacific Northwest National Laboratory, Richland, WA 99352, USA}
\author{A.~Jamil}\affiliation{Wright Laboratory, Department of Physics, Yale University, New Haven, CT 06511, USA}
\author{S.H. Kim}\affiliation{Lawrence Livermore National Laboratory, Livermore, CA 94550, USA}
\author{R.F.~Lang}\affiliation{Department of Physics and Astronomy, Purdue University, West Lafayette, IN 47907, USA}
\author{K.G.~Leach}\affiliation{Department of Physics, Colorado School of Mines, Golden, CO 80401, USA}
\author{B.G.~Lenardo}\affiliation{Physics Department, Stanford University, Stanford, CA 94305, USA}
\author{W.H. Lippincott}\affiliation{University of California, Santa Barbara, Department of Physics, Santa Barbara, CA 93106, USA}
\author{A.~Marino}\affiliation{Department of Physics, Colorado School of Mines, Golden, CO 80401, USA}
\author{D.N.~McKinsey}\affiliation{University of California, Berkeley, Department of Physics, Berkeley, CA 94720, USA}\affiliation{Lawrence Berkeley National Laboratory (LBNL), Berkeley, CA 94720, USA}
\author{E.H.~Miller}\affiliation{SLAC National Accelerator Laboratory, Menlo Park, CA 94025, USA}\affiliation{Kavli Institute for Particle Astrophysics and Cosmology, Stanford University, Stanford, CA 94305 USA}
\author{D.C.~Moore}\email{david.c.moore@yale.edu}\affiliation{Wright Laboratory, Department of Physics, Yale University, New Haven, CT 06511, USA}
\author{B.~Mong}\affiliation{SLAC National Accelerator Laboratory, Menlo Park, CA 94025, USA}
\author{B.~Monreal}\affiliation{Department of Physics, Case Western Reserve University, Cleveland, OH 44106, USA}
\author{M.E.~Monzani}\affiliation{SLAC National Accelerator Laboratory, Menlo Park, CA 94025, USA}\affiliation{Kavli Institute for Particle Astrophysics and Cosmology, Stanford University, Stanford, CA 94305 USA}
\author{I.~Olcina}\affiliation{Lawrence Berkeley National Laboratory (LBNL), Berkeley, CA 94720, USA}\affiliation{University of California, Berkeley, Department of Physics, Berkeley, CA 94720, USA}
\author{J.L.~Orrell}\affiliation{Pacific Northwest National Laboratory, Richland, WA 99352, USA}
\author{S. Pang}\affiliation{Lawrence Livermore National Laboratory, Livermore, CA 94550, USA}
\author{A.~Pocar}\affiliation{Amherst Center for Fundamental Interactions and Physics Department, University of Massachusetts, Amherst, MA 01003, USA}
\author{P.C.~Rowson}\affiliation{SLAC National Accelerator Laboratory, Menlo Park, CA 94025, USA}
\author{R.~Saldanha}\affiliation{Pacific Northwest National Laboratory, Richland, WA 99352, USA}
\author{S.~Sangiorgio}\affiliation{Lawrence Livermore National Laboratory, Livermore, CA 94550, USA}
\author{C.~Stanford}\affiliation{Department of Physics, Harvard University, Cambridge, Massachusetts, USA}
\author{A. Visser}\affiliation{Lawrence Livermore National Laboratory, Livermore, CA 94550, USA}

\begin{abstract}
Large detectors employing xenon are a leading technology in existing and planned searches for new physics, including searches for neutrinoless double beta decay ($0\nu\beta\beta$) and dark matter. While upcoming detectors will employ target masses of a ton or more, further extending gas or liquid phase Xe detectors to the kton scale would enable extremely sensitive next-generation searches for rare phenomena. The key challenge to extending this technology to detectors well beyond the ton-scale is the acquisition of the Xe itself. We describe the motivation for extending Xe time projection chambers (TPCs) to the kton scale and possible avenues for Xe acquisition that avoid existing supply chains. If acquisition of Xe in the required quantities is successful, kton-scale detectors of this type could enable a new generation of experiments, including searches for $0\nu\beta\beta$ at half-life sensitivities as long as $10^{30}$~yr.
\end{abstract}

\maketitle
\section{Introduction}
In recent years, detectors employing xenon have found applications in a variety of areas in nuclear and particle physics~\cite{Aprile:2009dv_review}. As a noble gas, Xe can be purified to extremely high levels, providing a high quality detection medium for ionization or scintillation light. In addition, this high purity allows Xe to serve as an ultra-low background material for rare event searches. Xe can also be liquefied at relatively high temperature (approximately 165~K at atmospheric pressure) and its high atomic number and density lead to higher stopping power for radiation than lighter gases such as He, Ne, or Ar. When incorporated into a homogeneous detector, this high stopping power allows Xe detectors to be compact, and effectively shields the inner regions of the detector from external radiation. 

The above properties have made Xe-based detectors among the most sensitive methods for searching for Weakly Interacting Massive Particles (WIMPs)~\cite{XENON:2018voc_ton_year,LUX:2016ggv_complete,PandaX-4T:2021bab}, neutrinoless double beta decay (\znu)~\cite{EXO-200:2019rkq_complete,KamLAND-Zen:2016pfg}, coherent elastic neutrino-nucleus Scattering (CE$\nu$NS)~\cite{RED-100:2019rpf_cevns}, and other rare phenomena including charged lepton flavor violation (cLFV)~\cite{MEGII:2018kmf}. Beyond these applications in fundamental physics, these properties also make Xe an appealing choice for compact radiation detectors in medical applications, such as Positron Emission Tomography (PET)~\cite{1976MedPh...3..283L_PET,Ferrario:2017sgq_PET}.

Despite these advantages, a key drawback to the use of Xe in large detectors is its high cost relative to lighter noble gases, and the limited quantities in which it can be obtained (see Sec.~\ref{sec:air_capture}). While the current market price and availability of Xe is possible because of large air liquefaction for the steel industry, this also leads to a relatively inelastic supply. The resulting price volatility and the supply shock inherent in a large purchase of Xe for scientific uses limits the feasible size of Xe detectors based on this supply chain to several 10s of tons. Existing and planned detectors are already reaching this scale.

However, it may be possible to develop alternative production methods for Xe that would avoid existing constraints, removing the fixed ceiling on current production and possibly also lowering acquisition costs. Here we describe ideas for methods of Xe acquisition beyond those employed by the fundamental science community to date, which may enable extremely large detectors. If Xe could be acquired in kton (kt) scale quantities at cost substantially below the current market price, it is plausible that Xe detectors could continued to be scaled to substantially higher masses. In particular, scaling Xe detectors to the kton scale may enable searches for \znu over the vast majority of allowed parameter space for the decay, searches for dark matter at larger scale than otherwise possible, measurements of solar $\nu$ that are complementary to existing techniques, and other extremely sensitive searches for new physics.

In the following sections we briefly describe the scientific motivation for kton-scale Xe detectors (Sec.~\ref{sec:motivation}), ideas for acquisition of kt quantities of Xe (Sec.~\ref{sec:air_capture}), and describe concepts for gas or liquid phase kton Xe time projection chambers (TPCs) that could reach \znu half-life sensitivities as long as $10^{30}$~yr (Sec.~\ref{sec:concepts}).

\section{Motivation}
\label{sec:motivation}
\subsection{Search for $0\nu\beta\beta$ decay in $^{136}$Xe}

Searches for \znu---in which an even-$A$ nucleus decays via emission of two $\beta$ particles, but no neutrinos---are uniquely sensitive to a number of Beyond-the-Standard-Model (BSM) physics scenarios. Recent community studies have placed high priority on further development of sensitive searches for \znu (see, e.g., Refs~\cite{Aprahamian:2015qub_NSAC_2015,HEPAPSubcommittee:2014bsm,Giuliani:2019uno_APPEC}), since observation of this decay would have far reaching consequences for fundamental physics. Regardless of the decay mechanism, observation of \znu would demonstrate that neutrinos are Majorana fermions~\cite{PhysRevD.25.2951_Schechter_Valle}. Discovery of Majorana neutrinos would confirm that a fundamentally new mass mechanism is realized in nature, which differs from that responsible for the charged fermion masses. In addition, if neutrinos do have Majorana masses, then lepton number violation (LNV) must occur. While both lepton number and baryon number are conserved in the Standard Model (SM) itself, the generation of the matter-antimatter asymmetry in the early universe requires extensions to the SM that violate conservation of baryon number, which possibly originate from LNV processes~\cite{Davidson:2008bu_leptogenesis}. Searches for LNV and the origin of neutrino mass are thus tightly entwined, and may have implications for fundamental open questions in cosmology.

Due to this motivation, a number of existing searches have been performed for \znu with isotope masses of $\sim$0.1~t, reaching half-life sensitivities between $10^{25}--10^{26}$~yr~\cite{EXO-200:2019rkq_complete,KamLAND-Zen:2016pfg,GERDA:2020xhi_final,PhysRevC.100.025501_MJD,CUORE:2021gpk_tonne}. Planned searches at the ton-scale aim to reach $\sim$10$^{28}$~yr sensitivity in the coming decade~\cite{nEXOpCDR:2018,LEGEND:2021bnm_pcdr,CUPIDpCDR:2019,NEXT:2020amj_NEXTtonne}. While the discovery potential of these ton-scale searches is significant, it is possible that \znu occurs at half-lives beyond the reach of ton-scale experiments. In this case, detectors at the kton-scale may be required to probe the majority of remaining parameter space for the decay (see Sec.~\ref{sec:0v_param_space}).

The key challenge to observe \znu is the extremely long half-life possible for the process. The half-life is related to the neutrino mass as:
\begin{eqnarray}
\left(T_{1/2}^{0\nu}\right)^{-1}=G^{0\nu}g_A^4|M^{0\nu}|^2\frac{\langle m_{\beta\beta}\rangle^2}{m_e^2},
\label{eqn:znudecayrate}
\end{eqnarray}
where $G$ is the two-body phase-space factor, $M$ is the nuclear matrix element (NME), $g_A$ is the axial vector coupling constant, and $m_e$ is the electron mass. The effective Majorana mass $\langle m_{\beta\beta}\rangle$ is a linear combination of the masses of the neutrinos ($m_j$ for $j=1,2,3$) that depends on the mixing angles measured in neutrino oscillation experiments, and on two unknown Majorana phases, $\alpha_1$ and $\alpha_2$~\cite{ParticleDataGroup:2020ssz,Benato2015}. Typical values for $\langle m_{\beta\beta}\rangle$ and $T_{1/2}^{0\nu}$ given current experimental constraints are described in Sec.~\ref{sec:0v_param_space}.

\begin{figure*}[t]
    \centering
    \includegraphics[width=\textwidth]{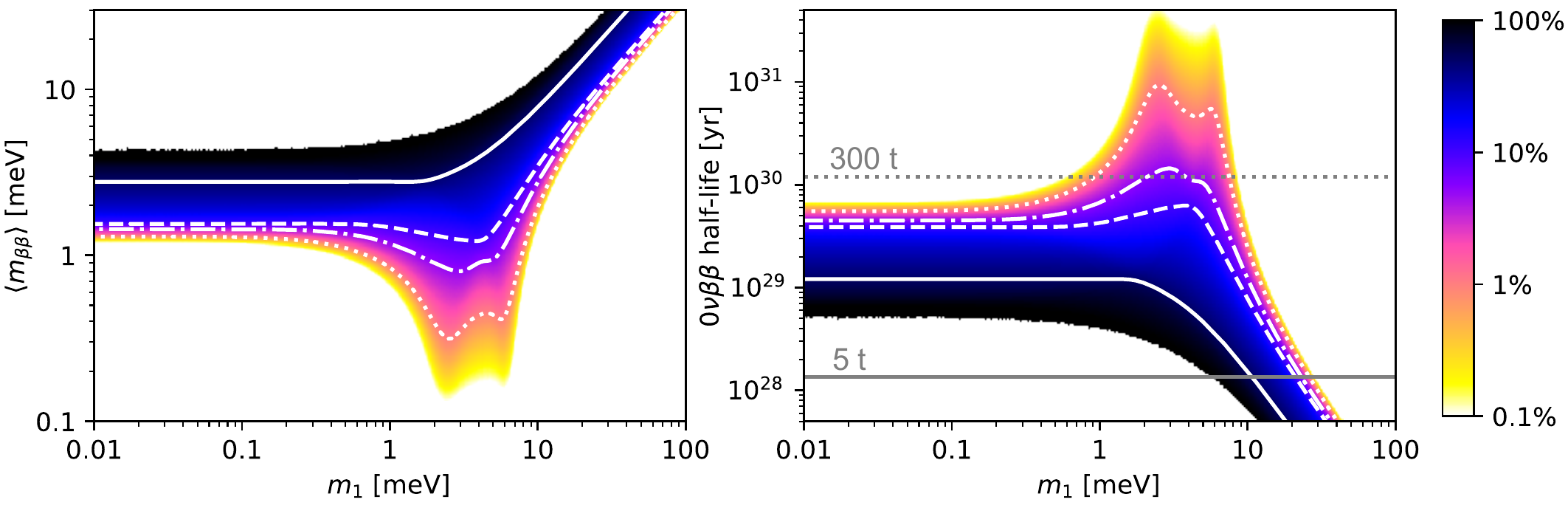}
    \caption{(left) Parameter space for the effective Majorana mass, $\langle m_{\beta\beta} \rangle$, in the normal ordering, as a function of the lightest neutrino mass, $m_1$. The inverted ordering is expected to be fully covered by planned ton-scale experiments, and the corresponding parameter space is not shown. At each value of $m_1$, the color scale indicates the probability for which $\langle m_{\beta\beta} \rangle$ is above a given mass assuming a uniform distribution for the unknown Majorana phases. The white contours indicate the sensitivity for which 50\% (solid), 90\% (dashed), 95\% (dash-dotted), and 99\% (dotted) of sampled values for $\langle m_{\beta\beta} \rangle$ lie above the curve at each value of $m_1$. (right) Conversion of the $\langle m_{\beta\beta} \rangle$ parameter space to half-life, assuming $|\mathcal{M}^{0\nu}| = 2.7$, which corresponds to the median value among currently published models~\cite{nEXO:2021ujk_nexo_new_sens}. The reach of a planned ton-scale detector containing a mass of approximately $5$~t of \isoXe~\cite{nEXO:2021ujk_nexo_new_sens} and a kton-scale detector with approximately $300$~t of \isoXe (see Sec.~\ref{sec:concepts}) are indicated.}
    \label{fig:majoranamass}
\end{figure*}

Two isotopes of xenon, $A=134$ and $136$, satisfy the conditions for undergoing $\beta\beta$ decay, with $^{136}$Xe being the most attractive for \znu decay searches given its relatively large $\beta\beta$-decay $Q$-value ($Q_{\beta\beta}=2458.10(31)$~keV)~\cite{PhysRevC.82.024603_Qval,PhysRevLett.98.053003_Qval} and natural abundance of 8.9\%~\cite{deLaeter_nat_abund}. Due to the lower $Q$-value of $^{134}$Xe ($825.8[9]$~keV)~\cite{Ame2012_134} and the expected $Q^5$ scaling in the decay rate~\cite{PhysRevD.84.093004_Duerr}, \znu decays from $^{134}$Xe are expected to be sub-dominant even in detectors employing Xe not enriched in $^{136}$Xe.
Existing or upcoming detectors searching for the \znu decay of $^{136}$Xe include gas-phase (e.g. NEXT-100~\cite{NEXT:2015wlq_next100_sens} and PANDAX-III~\cite{Chen:2016qcd_PandaXIII}) and liquid-phase (e.g. EXO-200~\cite{EXO-200:2019rkq_complete}) time projection chambers (TPCs), as well as liquid scintillator detectors (e.g. KamLAND-Zen~\cite{KamLAND-Zen:2016pfg}). The most sensitive searches to date employing Xe are EXO-200 and KamLAND-Zen, which set lower limits for the decay of \isoXe of $T_{1/2}^{0\nu\beta\beta} > 3.5 \times 10^{25}$~yr~\cite{EXO-200:2019rkq_complete} and $T_{1/2}^{0\nu\beta\beta} > 1.1 \times 10^{26}$~yr~\cite{KamLAND-Zen:2016pfg}, respectively. Planned ton-scale searches to be built in the coming years such as nEXO~\cite{nEXO:2021ujk_nexo_new_sens} and NEXT-1t~\cite{NEXT:2020amj_NEXTtonne} aim to search for \znu of \isoXe with a half-life sensitivity $\gtrsim 10^{28}$~yr.

\subsubsection{Parameter space for standard decay mechanisms}
\label{sec:0v_param_space}
If the effective Majorana mass \meff~$> 15$~meV, then the upcoming generation of ton-scale experiments will most likely discover \znu. This mass sensitivity corresponds to the full parameter space allowed in the inverted ordering, as well as a portion of the parameter space in the normal ordering where the lightest neutrino mass $m_1 \gtrsim 20$~meV. However, if the mass ordering is normal and $m_1 \lesssim 10$~meV, then \znu may be out of reach of planned ton-scale detectors even if neutrinos are Majorana particles. In this case, a larger detector would be required to explore the remaining allowed parameter space for the decay. 

Figure~\ref{fig:majoranamass} shows the allowed parameter space for \znu assuming the normal ordering, using current global fits to neutrino oscillation data (Nu-Fit v5.0~\cite{Esteban:2020cvm_nu_fit,nufit}). For each possible value of $m_1$, the allowed parameter space is calculated assuming that the two unknown Majorana phases are uniformly distributed on $[0,2\pi]$, following a similar methodology as Ref.~\cite{Benato2015}. The color scale in Fig.~\ref{fig:majoranamass} shows the probability at each value of $m_1$  that \meff would fall below a given sensitivity, under the above assumption for the unknown phases (and including uncertainties from the global fits to oscillation data). For $m_1 \lesssim 1$~meV or $m_1 \gtrsim 10$~meV, a detector reaching sensitivity of \meff~$\sim 1$~meV would fully probe the allowed parameter space for \znu. In the intermediate region with 1~meV~$\lesssim m_1 \lesssim 10$~meV, cancellations driving \meff below 1~meV are in principle possible for certain values of $\alpha_1$ and $\alpha_2$. However, assuming {\em a priori} that these phases are uniformly distributed, there is $\lesssim 5$\% probability that such a cancellation would occur at any of the values of $m_1$ in this range. Thus, for all values of $m_1$ possible in the normal ordering, a detector reaching such sensitivity would explore the vast majority of the allowed parameter space for \znu.

Searches for \znu directly constrain the decay half-life, which can be related to \meff through Eq.~\ref{eqn:znudecayrate}. However, significant uncertainties in this conversion arise from the theoretical uncertainty in the value of the NME~\cite{Dolinski:2019nrj_review}. As a benchmark for reaching 1~meV sensitivity on \meff, here we consider the Majorana mass reach of a hypothetical detector with a given half-life sensitivity, assuming a value for the NME corresponding to the median model among published results (see, e.g., the compilation of NME models in Ref.~\cite{nEXO:2021ujk_nexo_new_sens}). The phase space factor from Ref.~\cite{Kotila:2012zza} and $g_A = 1.27$ are assumed. For these values, a detector reaching $\gtrsim 10^{30}$~yr sensitivity would reach sensitivity corresponding to the \meff~$\lesssim 1$~meV benchmark, as shown in Fig.~\ref{fig:majoranamass}~(right).  While this represents the reach assuming the median NME model, the full range of NMEs published to date correspond to a sensitivity range for \meff between 0.6--2.5~meV at $10^{30}$~yr half-life sensitivity~\cite{nEXO:2021ujk_nexo_new_sens}.

At a \znu half-life of $10^{30}$~yr, the expected number of decays in a sensitive mass $m_{136}$ is
\begin{align}
\begin{split}
    R =&\ 0.3\ \mathrm{decays/yr} \left( \frac{m_{136}}{100\ \mathrm{t}} \right)\left( \frac{10^{30}\ \mathrm{yr}}{T_{1/2}} \right)\\
    =&\ 2.3\ \mathrm{decays/(kt\ yr\ FWHM)}\left( \frac{10^{30}\ \mathrm{yr}}{T_{1/2}} \right) .
\end{split}
    \label{eq:0v_rate}
\end{align}
Here, the sensitive mass is defined as the product of the \znu event detection efficiency, $\varepsilon$, the total detector mass, $m_{\text{det}}$, and the fraction of the detector mass consisting of \isoXe, $\eta$, such that $m_{136} = \varepsilon\, \eta\, m_{\text{det}}$. For \natXe, $\eta = 0.089$, while an enriched detector could have an isotope fraction as large as $\eta =$ 0.8--0.9~\cite{nEXOpCDR:2018,NEXT:2020amj_NEXTtonne,Auger:2012gs_exo200jinst,KamLAND-Zen:2016pfg}.
In the second line of Eq.~\ref{eq:0v_rate}, the mass in kt corresponds to $m_{136}$, and converting to detector mass would require scaling by $\varepsilon$ or $\eta$ if either differs from unity.
The above rate approximately indicates that a total exposure of $\gtrsim 1$~kt~yr is required at a half-life sensitivity $\sim$10$^{30}$~yr, if a perfectly-efficient, background free detector could be constructed. Based on the detector concepts considered in Sec.~\ref{sec:detector_concepts}, a practical detector would require slightly higher quantities of Xe to reach this sensitivity, i.e. approximately 0.3~kt of \isoXe or 3~kt of \natXe. As described in Sec.~\ref{sec:air_capture}, new ideas would be required to acquire Xe in sufficient quantities for such a detector.  

\subsubsection{\label{mflnv}Majorana Fermions and LNV}
While the above discussion focuses on the standard decay mechanism, alternative extensions to the SM generating \znu have been studied, in many cases with substantially enhanced decay rates (see, e.g., Ref.~\cite{Dolinski:2019nrj_review} for a recent review). Regardless of the decay mechanism, searches for \znu will remain among the most powerful generic probes for LNV in the coming decades, with significant complementarity to other precision tests~\cite{PhysRevD.95.115010_LNV}. Extending the reach of such searches to half-life sensitivities as long as $10^{30}$~yr would thus allow more than two orders-of-magnitude extension in parameter space for LNV processes, beyond existing and planned experiments. 

\begin{figure}[t]
    \centering
    \includegraphics[width=\columnwidth]{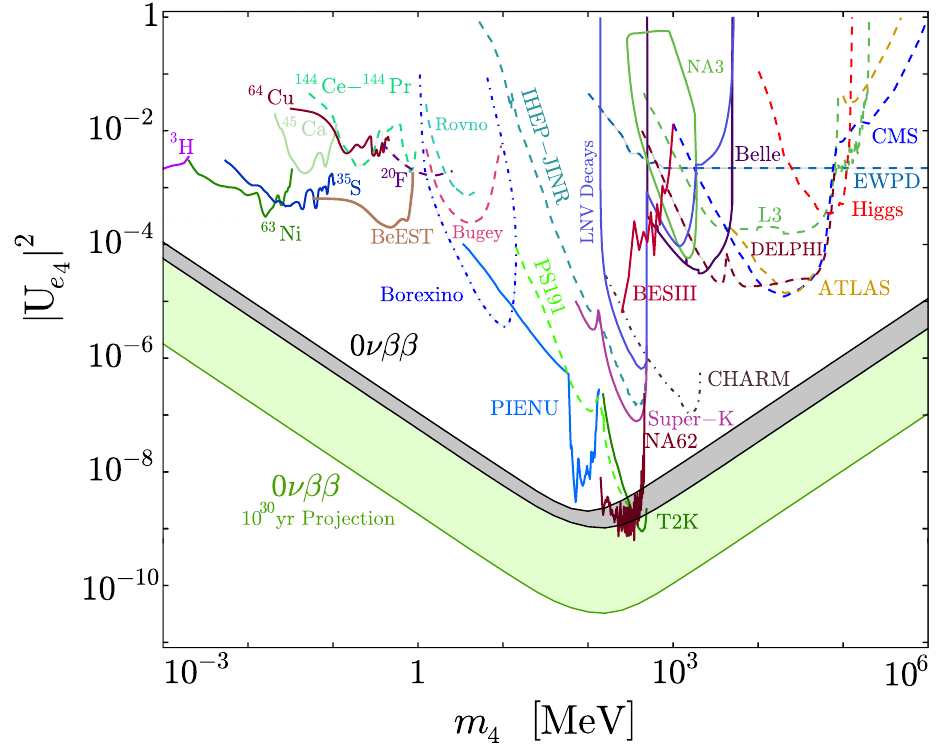}
    \caption{Constraints on the flavor mixing strength $|U_{e4}|^2$ of a single sterile neutrino with the electron flavor as a function of its mass $m_4$. The curves represent the current limits from experiments as labeled, with varying levels of model dependencies. The ``$0\nu\beta\beta$'' band denotes the current limit from \znu decay searches with a Majorana sterile neutrino from Ref.~\cite{PhysRevD.103.055019} where the band is the uncertainty due to nuclear matrix elements.  The projected limits from a kt-scale \znu decay search in $^{136}$Xe are also shown, where the light shaded region highlights the nearly two order of magnitude improvement in sensitivity.  Figure adapted from Ref.~\cite{PhysRevD.103.055019} with updates from Ref.~\cite{PhysRevLett.126.021803}.}
    \label{fig:0vsterilelimits}
\end{figure}

In an effective field theory approach, augmenting the SM Lagrangian with operators with mass dimension $>4$ can introduce LNV phenomena. The lowest dimension operator of this type, a dimension-5 operator known as the ``Weinberg operator''~\cite{PhysRevLett.43.1566_Weinberg,Rodejohann:2011mu_0vreview}, can introduce LNV associated with the corresponding effective energy scale for the operator, $\Lambda$. Existing searches can probe effective scales $\Lambda \approx 10^{11}$~TeV~\cite{PhysRevD.95.115010_LNV, Rodejohann:2011mu_0vreview}, corresponding to a sensitivity to the neutrino mass $\approx 100$~meV. A search at $10^{30}$~yr half-life sensitivity would reach effective neutrino mass scales of $\approx 1$~meV, corresponding to $\Lambda \approx 10^{13}$~TeV, i.e. the GUT scale~\cite{ParticleDataGroup:2020ssz}. Thus, searches for \znu represent one of the only known laboratory techniques for accessing possible new phenomena at such high energies (searches for $p$ decay also present another important parallel path to phenomena at this scale~\cite{Super-Kamiokande:2014otb_p_decay}). If higher dimension operators are considered, \znu remains among the most sensitive generic probes for LNV, complemented by parallel searches for flavor-violating processes~\cite{PhysRevD.95.115010_LNV}.

\subsubsection{Heavy Neutral Leptons and Massive Scalar Emission}

 As a concrete example of a general class of models beyond the standard mechanism, the addition of sterile neutrinos in extensions to the SM can substantially modify the \znu decay rate.
In the simplest case, considering a single sterile neutrino with mass $m_4$ and neglecting the contribution from the active neutrinos, the current non-observation of \znu decay can produce significant constraints on the presence of such sterile $\nu$ over a wide mass range~\cite{PhysRevD.103.055019,PhysRevLett.126.021803}, as shown in Fig.~\ref{fig:0vsterilelimits}. 

As described above, such mechanisms may allow discovery of \znu if, e.g., they substantially enhance the rate relative to the standard decay mechanism described in Sec.~\ref{sec:0v_param_space}. Alternatively, the absence of an observation of \znu at half-lives up to $10^{30}$~yr could place further constraints on the presence of such sterile $\nu$. However, some caveats apply to these exclusions. If the active and sterile neutrinos are purely Dirac fermions, lepton number cannot be violated through this mechanism and thus \znu decay is forbidden. Further, since the heavy and light mass states are connected via the seesaw relation, if the sterile states are lighter than the \znu decay momentum transfer,
the \znu decay rate will be suppressed. More extensive discussions on the relation between \znu decay and sterile neutrinos are included in Refs.~\cite{Bolton2020,PhysRevD.103.055019,Abada2019,Dekens2020}.

\subsection{Other possible applications}
While in this work we primarily focus on motivations for kton-scale Xe detectors for searches for \znu and LNV, here we briefly highlight additional applications that may be possible with such detectors. A multipurpose detector, e.g., optimized for searches for \znu, dark matter, and possibly measurements of solar or supernova $\nu$ may be possible, although further study of tradeoffs between different applications would be required. Regardless of the ultimate optimization between dedicated and multipurpose detectors, the ideas for Xe acquisition described here may enable a new generation of detectors for a variety of rare event searches beyond \znu.
\subsubsection{WIMPs}
\label{sec:DM}
There is now overwhelming astrophysical evidence that dark matter constitutes a majority of the matter in the Universe~\cite{ParticleDataGroup:2020ssz}, but its nature has yet to be understood. 
Weakly interacting massive particles (WIMPs)~\cite{PhysRevD.31.3059_goodman} are a well-motivated class of dark matter candidates, and
LXe TPCs are currently the leading technology to search for WIMPs in terrestrial detectors from masses of $\sim$3~GeV/c$^2$ to several TeV/c$^2$~\cite{LUX:2016ggv_complete,XENON:2018voc_ton_year,PandaX-4T:2021bab}. Recent results from a 1~t\,yr exposure of LXe set a 90\% CL upper limit on the WIMP-nucleon spin-independent elastic scatter cross-section at $5 \times 10^{-47}$~cm$^2$ for a 50~GeV/c$^2$ WIMP~\cite{XENON:2018voc_ton_year}, approximately two orders-of-magnitude better than current limits from technologies other than LXe TPCs at this mass. Data taking with $\sim$6--7 ton liquid xenon TPCs is currently underway with a projected sensitivity of roughly $1.5 \times 10^{-48}$~cm$^2$ for a 50~GeV/c$^2$ mass WIMP~\cite{XENON:2020kmp_xenonnt,Akerib:2018lyp}. Additionally, a future 40~t detector with a total exposure of 200~t\,yr aims to extend sensitivity down to $2.5 \times 10^{-49}$~cm$^2$ at the same mass~\cite{DARWIN:2016}. A practical constraint on the sensitivity for such WIMP searches arises from the atmospheric neutrino background. CE$\nu$NS of atmospheric neutrinos with Xe is indistinguishable on an event-by-event basis from the WIMP signal in LXe TPCs, and hence sensitivity to WIMPs is limited by the systematic uncertainty on the atmospheric neutrino background rate. Assuming a roughly 20\% systematic uncertainty on the atmospheric neutrino flux, at 50 GeV this so-called neutrino ``fog" or ``floor" corresponds to a cross-section of approximately $10^{-49}$ cm$^2$ ~\cite{Billard:2021uyg_appec} and a total xenon exposure on the order of a~kt\,yr. 

Reaching sensitivities approaching the neutrino floor appears to be achievable with extensions to existing technologies~\cite{DARWIN:2016} and with existing Xe supply chains. If WIMPs are discovered near the $\nu$ floor, larger detectors may be needed to study their properties in detail. In the absence of such a discovery, scaling such detectors to the kt scale (due to the strong motivation from, e.g., \znu searches) would allow further high-sensitivity searches for WIMPs, possibly with a multi-purpose detector. While CE$\nu$NS and WIMP scattering have the same event-by-event signature, statistical separation is in principle possible with large numbers of events, e.g. through the expected annual modulation of the WIMP scattering event rate (although this would have to be carefully separated from the similar known annual modulation of atmospheric muon production)~\cite{Freese:2012xd}. For a detector sensitive to the direction of the recoil, the expected diurnal modulation in the direction of WIMP recoils and atmospheric $\nu$ could be separated~\cite{Mayet:2016zxu}. Such directional sensitivity might in principle be possible in GXe TPCs~\cite{Nygren_2013}, but has not yet been fully demonstrated.

\subsubsection{Alternative dark matter models}
Given the lack of detection to date of WIMPs (or other highly motivated candidates such as axions~\cite{Graham:2015ouw_axions}), a large number of alternative models have been studied (see, e.g. Ref.~\cite{ParticleDataGroup:2020ssz}). For general classes of models where dark matter (or some sub-component of the relic density) consists of much heavier particles than typical WIMPs ($\gg$TeV, including composite particles~\cite{PhysRevLett.64.615_unitarity}), these particles could have evaded detection to date due to their relatively low flux through existing meter-scale detectors. An extremely large LXe or GXe TPC could identify such dark matter candidates if they produce energy depositions in the keV--MeV range, below the threshold, e.g., of other kton-scale liquid scintillator $\nu$ detectors.

In addition, a variety of models have been studied for dark matter that primarily produce energy depositions in the MeV range, for either electron or nuclear recoils~\cite{An:2012bs,Dror:2019dib,Bringmann:2018cvk}. Searches for several such dark matter candidates have been performed by existing detectors originally designed for $\nu$ physics (see, e.g.,~\cite{PhysRevLett.120.211804_mcp,Majorana:2016hop,Abe:2021nem,PROSPECT:2021awi}), and further scaling these searches to kton-scale masses would typically provide several orders-of-magnitude additional sensitivity.

\subsubsection{Neutrino Detection}
\label{sec:nu_detection}
Direct detection of neutrino interactions in a kton scale Xe TPC is also expected to be possible. Detectable interactions include coherent nuclear scattering (i.e., CE$\nu$NS) from Xe nuclei of atmospheric $\nu$ at keV energies, as well as elastic scattering (ES) of solar $\nu$ from electrons at MeV energies. These interactions primarily lead to backgrounds for WIMP searches and \znu, respectively, rather than signals by themselves. However, supernova neutrinos may also be detectable through these signatures if a sufficiently close supernova were to occur during detector operations. The sensitivity to such supernova $\nu$ for a 40~t TPC has recently been evaluated~\cite{Lang:2016zhv_super,XMASS:2016cmy_super}. A kton-scale TPC with sufficiently low threshold to observe CE$\nu$NS would further increase the distance and mass range over which such a burst could be detected. Due to its sensitivity to supernova neutrinos of all flavors, detection of supernovae $\nu$ through CE$\nu$NS would provide complementary information to other larger scale neutrino detectors observing such a burst~\cite{Scholberg:2012id_review}. 

Charged-current (CC) interactions of solar $\nu$ in a kton Xe TPC are also detectable. The unique signature of such interactions (including multiple de-excitation $\gamma$s from the excited $^{136}$Cs daughter nucleus, and its subsequent $\beta$ decay) allows their tagging and removal as backgrounds in the rare-event searches above (see Sec.~\ref{sec:solar_nu_bkgs}). However, this signature may also have the potential for background free identification of solar $\nu$ interactions via a delayed coincidence in Xe TPCs, if intermediate nuclear states are sufficiently long lived~\cite{PhysRevD.102.072009_cc_solar}. Detection of such solar $\nu$, including precise measurements of CNO $\nu$ or the $^{7}$Be lineshape could provide constraints on solar models that are complementary to existing measurements~\cite{PhysRevD.102.072009_cc_solar}. While such signatures may already be potentially detectable in ton-scale experiments, extensions of Xe TPCs to the kton scale would substantially enhance the statistical accuracy of such measurements.

\section{Xenon Extraction from Air}
\label{sec:air_capture}

Based on Eq.~\ref{eq:0v_rate}, reaching the $10^{30}$~yr half-life sensitivity benchmark for \znu would require $\gtrsim$kton-scale quantities of Xe to be obtained (containing $\gtrsim$100~t quantities of \isoXe). As will be described in Sec.~\ref{sec:concepts}, extensions of existing detector technology to this scale are plausible, and therefore the production of the Xe itself is the key challenge to enable such searches for \znu. The following sections briefly summarize existing methods for Xe production and identify techniques that may provide a path to acquisition of kton-scale quantities of Xe.

\subsection{Summary of current Xe production}

Commercial Xe is produced by separation from the atmosphere, where it is present at a concentration of $87\pm 1$~nL/L~\cite{airliquide}.  The total mass of Xe in the atmosphere is approximately 2~Gtons (assuming the mass of atmosphere is $5.1\times10^{21}$ g~\cite{CRCHCP2021}) providing an ample supply from which Xe could in principle be obtained. Xe is also naturally present in ground water, and is produced in nuclear reactors, although we are not aware that extraction of Xe from either source has been commercialized to date. Development of processes to extract Xe from nuclear fuel reprocessing are underway, but are unlikely to produce enough Xe for the kton-scale detectors considered here (but, may be of interest for intermediate scale detectors, as described in Sec.~\ref{sec:alternatives}).

Cryogenic liquefaction followed by distillation is the current method used to extract Xe from the atmosphere. 
 The cost of the Xe produced in this process benefits from the synergistic production of other valuable products such as liquid oxygen produced for the steel industry. 
Xe and other rare gases are concentrated in the oxygen sump and are distilled to separate the Xe from the liquid oxygen streams.  The dependence of Xe production on the steel industry lowers the cost, but it also limits the total world's production of Xe to 50--100~t/yr~\cite{xe_prod_rep}.  Cost and availability are acceptable for current experiments at the ton-scale, but both become limiting at the kton scale using the current Xe production methods.

Increasing the supply of Xe produced by cryogenic liquefaction beyond that corresponding to the demand for liquid oxygen by the steel industry is not viable at the scale considered in this paper. However, any industry that already processes large amounts of air but does not currently collect xenon (such as air separation plants using either cryogenic or pressure swing adsorption) should be considered for the synergistic possibility of sharing the energy cost of air movement. There is also growing interest in separating CO$_2$ and water from the atmosphere \cite{direct2016perez}, and these processes, if practiced at an industrial scale, may enable the addition of xenon extraction and a sharing of the energy cost to move and process the air. 

The thermodynamic minimum energy to separate Xe from air is only 42.1kJ/mol~\cite{Downie2002}, corresponding to a fundamental lower limit to the cost to produce Xe $\gtrsim$\$0.01/kg (assuming an energy cost of \$0.10/kWh). While no practical process could approach this fundamental limit, it is approximately 5~orders of magnitude lower than the current wholesale cost of Xe, allowing the possibility at least in principle for lower cost production through other techniques. These simple estimates motivate the consideration of alternative techniques to cryogenic liquefaction described in the following sections.

\subsection{Possible alternative techniques}
The low concentration of Xe in the atmosphere requires processing extremely large quantities of air to separate significant quantities of Xe. The movement and even minimal compression of this airstream can be the major energy cost, leading to the high costs described previously. A variety of alternative techniques that could avoid this costly compression were considered.

Cryogenic techniques can directly cool the air to separate the Xe. To optimize the efficiency of such techniques, the energy used to cool the gas must be recovered with high efficiency via a heat exchanger that transfers heat from the input air stream to the output waste stream.  As the heat exchanger approaches 100\% efficiency, the cooling power requirement becomes negligible.  The primary challenge with this method is building a heat exchanger that is effective enough to accommodate the extremely large air flow, with low pressure drop, while maintaining an extraordinarily high efficiency. For example, processing of 218 million liters/hr of air flow is required to extract 1~t of Xe per year at 100\% efficiency. More sophisticated versions of this basic idea could employ cryogenically cooled activated charcoal to capture the Xe, allowing higher temperature operation, but still facing similar challenges related to developing a sufficiently high efficiency heat exchanger.

Non-cryogenic separation techniques are also possible, where Xe can be adsorbed by suitable materials directly from the air stream. In adsorptive processes, atoms are trapped on the surface of an adsorbent material, either due to physical or chemical bonding. The amount of adsorbate present on the surface of a given material depends on the process conditions, primarily partial pressure and temperature. 
Through changes in these parameters, it is possible to vary the concentration of adsorbate atoms in the output stream compared to the feed stream. 
Separation processes via adsorption have made significant advances in recent years due to the development of ultra-high surface area microporous materials, such as carbons, zeolites, metal-organic frameworks, etc.
Examples include oxygen concentrators \cite{Ackley2019MedicalOC}, CO$_2$ capture systems \cite{ChoiCO2Materials,CO2MOFYaghi}, and hydrogen storage \cite{DURBIN201314595, SculleyHydrogen}. 

Materials that selectively adsorb Xe have been recently developed and provide perhaps a more promising approach than cryogenic distillation \cite{Banerjee2016, Xiong2018}.
Extraction of small quantities of Xe from atmospheric air has been demonstrated using activated carbon and zeolites \cite{Byun:2020,CAGNIANT2018461,RINGBOM2003542}.  Modification and scale up of these systems might be possible, but they have already been optimized to some degree and it does not appear that they will likely be scaled for the extraction of large quantities of Xe. Metal-organic frameworks are particularly attractive as they can be engineered at the molecular level to match desired adsorption properties.

Beyond the adsorbent material itself, an adsorption cycle in which the Xe is first adsorbed on the material and then desorbed from its surface for collection is required. The most common method is pressure swing adsorption (PSA), in which input air is compressed to increase adsorption on the adsorbent, and once saturated, the pressure decreased to desorb the Xe. 
The energy requirement is likely still too high even in a well-optimized system.

Vacuum swing adsorption (VSA) provides another possible alternative. For VSA, the input air stream is not compressed and the adsorption happens at atmospheric pressure. Once saturated, the Xe is desorbed at vacuum pressures. Because the vacuum is only required for the much smaller Xe stream, and the overall input airstream avoids compression, the energy required can be substantially reduced relative to PSA. 

Finally, thermal swing adsorption (TSA) does not require any pressure variations. The air flows over the adsorbent at ambient pressure and temperature and the Xe is desorbed by raising the temperature of the bed. Since the energy used to heat the bed can be efficiently recovered and is a lower quality energy (in comparison to PSA or VSA, where recovering energy used to pressurize gases is more difficult), TSA can in principle operate at very high efficiency relative to other methods.

\subsection{R\&D for Xe Separation via TSA}

Based on the considerations above, we consider here a specific concept for Xe separation based on a TSA cycle employing a metal-organic framework (MOF) material. While demonstrating the full feasibility of such a concept is beyond the scope of this paper, and subject of ongoing R\&D, here we highlight the availability of the key components and the main aspects of the R\&D.

The key design factors that drive the energy efficiency and capital costs for the process are the specific pressure drop and the adsorbent properties.  Beyond these primary factors, there are a number of important engineering challenges that must be addressed for practical implementations, including: multi-bed systems, reflux, intensification, possible gas pre-processing for water or CO$_2$, and heating methods.  However, here we focus only on the two primary drivers above.

\begin{figure}[t]
    \centering
    \includegraphics[width=\columnwidth]{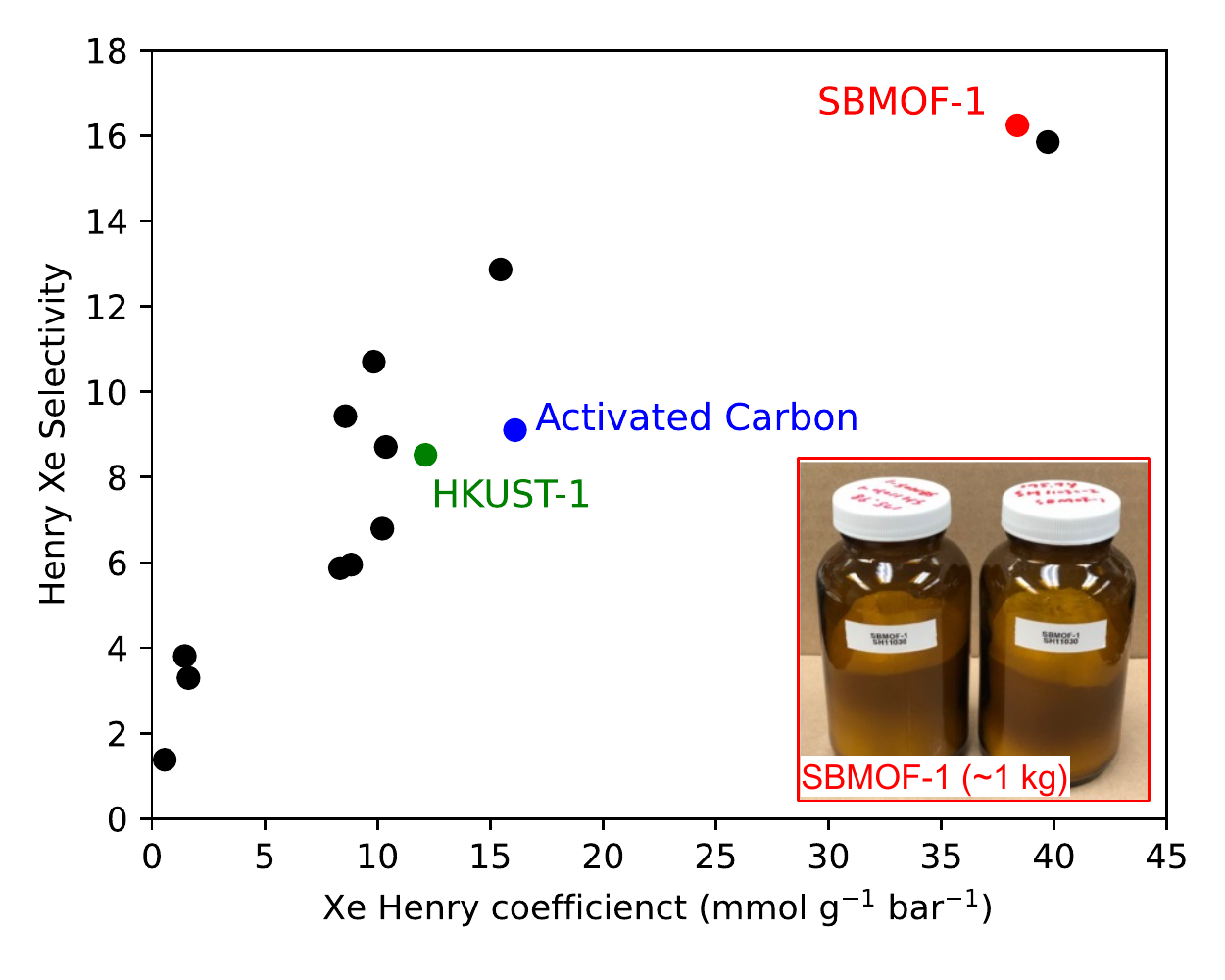}
    \caption{ Properties for several selected candidate materials that adsorb Xe, with colored points highlighting example materials discussed in the main text. The highest performing materials are in the upper right with high selectivity and Henry coefficient. Data compiled from from Refs.~\cite{Xiong2018,C8TA02091D_gong}. (inset) Production of SBMOF-1 in kg quantities from initial R\&D work. }
    \label{fig:sel-vs-henry}
\end{figure}

A significant amount of relevant work on materials for the separation of Xe (e.g.~\cite{Banerjee2016}) comes from work to separate Kr and Xe from the waste stream in nuclear reactor fuel reprocessing.  These studies provide measurements of the selectivity of the material (ratio of the adsorbed species divided by the ratio of the gas  partial pressures) and its Henry coefficient (ratio of the concentration of a species in the adsorbent and the gas phase at equilibrium), which is a measure of the affinity of the material for the adsorbate of interest. Figure~\ref{fig:sel-vs-henry} compares the performance of a number of materials.  The ability to cost-effectively synthesize the adsorbent in large quantities is also important, and has potential trade-offs with other parameters. For example, the HKUST-1 MOF has been produced in large quantities
and is relatively inexpensive, but does not have particularly high selectivity or affinity for Xe.  New MOFs, such as SBMOF-1, have been designed with tailored pore sizes to improve the selectivity and/or affinity for Xe, though are not yet commercially available.  A high-performing MOF like SBMOF-1 has already been synthesized at the $\sim$kg scale (Fig.~\ref{fig:sel-vs-henry}~[inset]), and a cost-effective scale up to larger quantities appears feasible through industrial partnerships.  Additional considerations for a given material are the specific adsorption capacity of the bed, stability of the material to other species in the gas mixture (e.g. water, oxygen), the adsorption kinetics, the selectivity to components of the air such as CO$_2$ and water, and optimizing the ratio of adsorbent to other thermal mass in the bed.  Previous work~\cite{Banerjee2016} and ongoing R\&D indicate that SBMOF-1 may have satisfactory properties, providing a starting point for investigation of Xe separation at large scales with these techniques.

The capital cost and energy efficiency can both be optimized by the choice of packing (i.e., geometrical arrangement) of the adsorbing material.  Typical beds consisting of a tightly packed, but random, arrangement of adsorbent beads are simple and cheap to manufacture, but suffer from high pressure drop and poor mass transfer kinetics.  Structured beds in which the adsorbent is arranged in a fully controlled geometry can be optimized to improve the performance by providing a smaller diffusion path, increasing the mass transfer, and lowering the specific pressure drop~\cite{Rezaei2010,AKHTAR20141643}.  Laminate adsorbent beds have been produced cost effectively for carbon capture from the air and are also well-suited for Xe capture~\cite{zeolitelaminate}. Methods to form the MOF into a structured adsorbent typically require a binder that does not damage the MOF, or hinder diffusion into the crystal, has a low heat capacity, and is not required in large mass fractions to bind the MOF. R\&D to date with SBMOF-1 has explored multiple avenues to build a bed and demonstrated that a laminate bed that meets the above requirements appears feasible, with additional studies ongoing.

Optimization of the process cycle and evaluation of its economic feasibility can be studied with simulations, including through industry-standard tools such as Aspentech aspenONE~\cite{aspen}. Preliminary simulations of a rapid-cycle TSA using a laminate structured adsorbent were performed with measured characteristics of SBMOF-1 as the adsorbent material  as inputs to the model. The cycle and the structured adsorbent parameters were adapted from an existing design for CO$_2$ sequestration. While preliminary, results of these simulations indicate that a pilot plant producing about 1~t/yr of Xe could generate Xe near the current production costs. Further improvement in the costs at larger scale is possible. In particular, an advantage of the TSA concept is the low quality energy required (i.e., low temperature heat and mechanical air movement), allowing many possible optimizations for a large scale plant.

Having identified these basic parameters for the TSA concept, an intermediate goal is to produce a small-scale prototype demonstrating Xe separation with a full cycle. The performance of such a prototype can be used to verify the accuracy of simulations of the system, which can then be scaled to project the performance of a larger pilot plant. Such a pilot plant is likely required to inform projections of the cost for Xe production in an optimized, full-scale plant.      

\subsection{Enrichment}

While detector concepts that do not require enrichment are considered in Sec.~\ref{sec:concepts}, in certain cases enrichment may be desirable for LXe detectors to suppress backgrounds from solar $\nu$ at the longest half-lives considered.
If enrichment is desired, centrifuge separation likely provides the preferred enrichment method due to its low operating cost and power requirements.  As an inert, noble gas, Xe is straightforward to separate in several centrifuge designs in current use today. Over a ton of \enrXe has already been produced and the current approximate cost is \$8--10/g, for production rates at the ton scale to 90\% enrichment.  The optimal enrichment level, taking into account costs, for a kton scale detector may be lower than at the ton scale (see Sec.~\ref{sec:concepts}), since enrichment at lower levels is less expensive. However, if enrichment is desired and costs are not substantially reduced relative to the ton scale, they might exceed the acquisition cost of the feedstock itself.

Overall enrichment costs require accounting for the capital construction costs, operation costs, and economic value of the enriched products and depleted tails. Centrifuge enrichment plants require larger capital costs than other technologies, and therefore extending the time to produce the Xe will likely have a large impact on the cost.  At the kton scale---and even at the few ton/yr scale---the enrichment capacity would have to be constructed, since no idle plants have sufficient capacity.  Assuming the supply chain for the centrifuge parts can support the required scale, a (likely conservative) cost estimate for enrichment at the kton scale would be to assume the current cost at the ton scale. Because the bulk of the cost is in the capital, the cost of the Xe could be substantially reduced if the production time can be extended. The enrichment costs could be offset, perhaps completely, by selling the depleted Xe.  Given the current cost of \natXe and the natural abundance of \isoXe, the depleted Xe is approximately of the same value as the \enrXe extracted for the experiment at current prices.  

A careful optimization of the cost and performance is required to determine if enrichment is needed, which is beyond the scope of the considerations here. Nonetheless, while expensive, enrichment at the required level may be feasible with existing technologies. Enrichment costs can be reduced by careful planning.

\section{kton-scale Xe TPC concepts for \znu searches}
\label{sec:concepts}
If Xe acquisition at the kton-scale is successful (Sec.~\ref{sec:air_capture}), scaling either liquid or gas Xe TPC technology to the kton scale is expected to be technologically feasible. Indeed, for liquid Ar TPCs where isotope acquisition issues are not dominant, experiments such as DUNE will employ multiple 17~kt TPCs in the coming years (with total active mass of 40~kt)~\cite{DUNE_TDR:2020}. As described below, the required scale for a Xe TPC to reach \znu half-lives as long as $10^{30}$~yrs is substantially more modest---roughly 3 kt of \natXe (or 300~t of \isoXe). While detector backgrounds are challenging for any \znu search at this scale, detectors reaching the required performance would primarily require scaling up already demonstrated techniques to larger sizes. In the following sections we review the primary backgrounds that influence the design of kton-scale Xe TPCs, concepts for gas and liquid phase detectors, and the advantages of such TPCs compared to other proposed detector technologies.

\subsection{Backgrounds}
\label{sec:backgrounds}
Based on Eq.~\ref{eq:0v_rate}, at $T_{1/2} \sim 10^{30}$~yr background rates $\lesssim$2~events/(kt~yr~FWHM) are required to give a signal-to-background ratio $\gtrsim$1. This represents a substantial reduction in background rate relative to the current generation of \znu detectors, which have projected effective backgrounds $\gtrsim 500$~events/(kt~yr~FWHM)~\cite{Agostini:2017jim,nEXO:2021ujk_nexo_new_sens,NEXT:2020amj_NEXTtonne,CUPIDpCDR:2019}. Homogeneous detectors such as the gas and liquid phase TPCs considered here may be able to reduce sources of external backgrounds that are dominant in ton-scale experiments simply by scaling to the kton-scale. For such detectors, other backgrounds are expected to become dominant, including those arising from the tail of the \tnu spectrum or from elastic scattering of solar $\nu$. 
\subsubsection{External backgrounds}
\label{sec:external_bkgs}
The dominant backgrounds in planned ton-scale detectors typically arise from external radiogenic backgrounds~\cite{nEXOpCDR:2018,LEGEND:2017,CUPIDpCDR:2019}. A key advantage of large, homogeneous liquid and gas phase detectors is the ability to purify the detector medium {\em in situ}, so that $\gamma$ backgrounds from natural U/Th radioactivity arise only from external sources, i.e., materials surrounding the Xe. At the kton-scale, it remains to be demonstrated that the Xe (or indeed any other possible detector material) can be purified to sufficiently remove non-noble gas radioactivity to the level that backgrounds from internal U/Th are negligible. However, the ability to recirculate and purify gas or liquid phase noble elements may provide a path to the required purity. Instead, U/Th-chain activity within the LXe is expected to be dominated by $^{222}$Rn emanation into the Xe (discussed separately in Sec.~\ref{sec:internal_bkgs}).

\begin{figure}[t]
    \centering
    \includegraphics[width=\columnwidth]{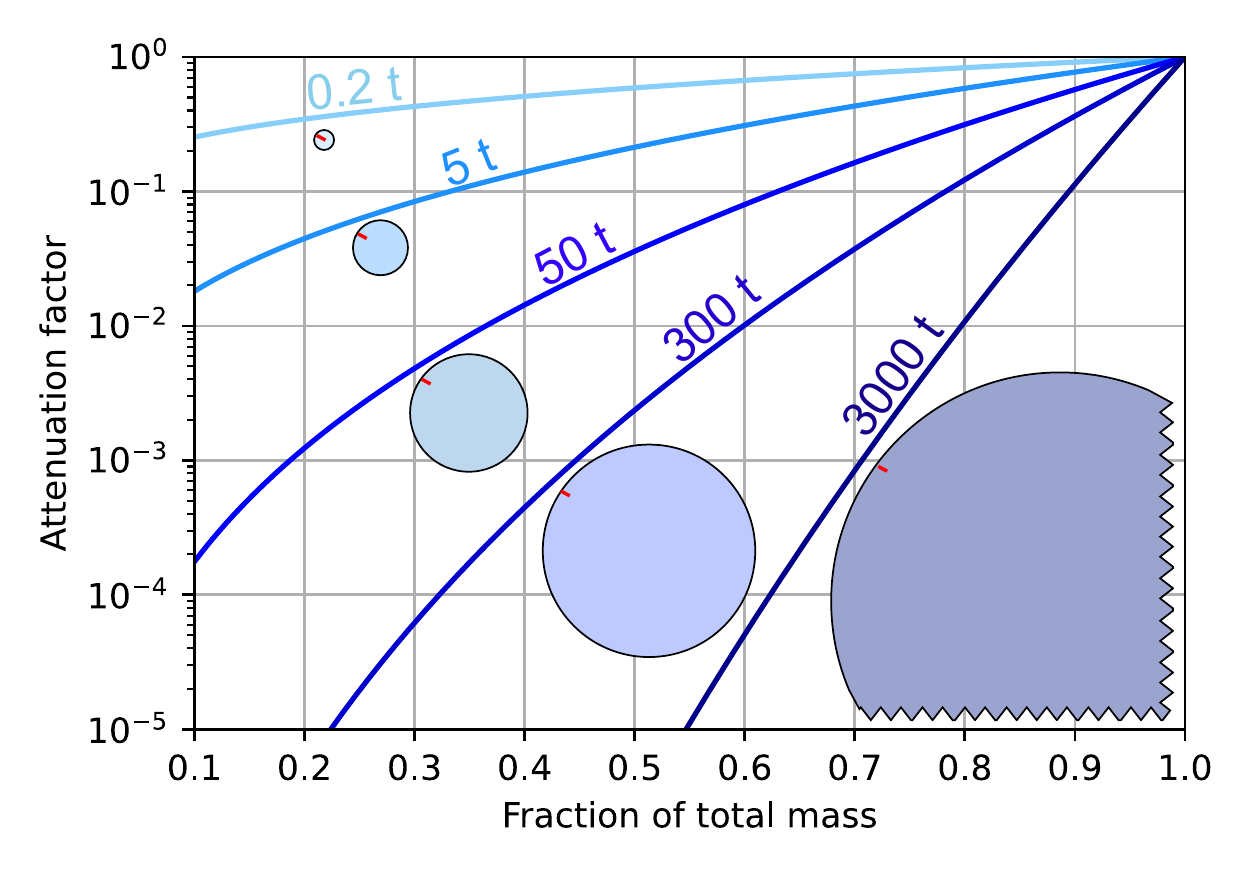}
    \caption{ Schematic of self-shielding from external backgrounds as LXe detectors are scaled to larger size. The attenuation factor, $e^{-d/\lambda}$, for a $\gamma$ traveling a distance $d$ into the detector versus the total mass beyond this distance from the walls is shown. The insets show corresponding cross-sections for a square cylinder of the given mass. The $\gamma$ attenuation length is $\lambda \approx 8.5$~cm (at 2.5~MeV), while for visibility in the plot, the red line indicates the distance for a factor of 10 attenuation (i.e., 2.3$\lambda$). }
    \label{fig:external_bkgs}
\end{figure}

External backgrounds arising from the surface of the detector are strongly attenuated by the ``self-shielding'' of the Xe, with mass attenuation coefficient $\mu/\rho = 0.038$~cm$^2$/g at 2.5~MeV~\cite{nist_xcom_148746}. This attenuation coefficient corresponds to a linear attenuation length of 8.5~cm for liquid Xe. For gas, the self-shielding is less effective (at the same total mass) due to the lower density, with the attenuation length varying between 2.6--0.5~m for GXe densities between 0.1--0.5~g/cm$^3$ (i.e., pressure between 15--50~bar). Since for both gas and liquid these attenuation lengths are small compared to the linear dimensions of a kton-scale Xe TPC, the rate of backgrounds arising from external sources is substantially reduced in the inner regions of the detector, as shown in Fig.~\ref{fig:external_bkgs}, for the LXe case. In addition, kton-scale detectors generally benefit from the reduced surface-to-volume ratio at larger size. Detailed quantification of this self-shielding of external backgrounds for example LXe and GXe detector concepts is described in Sec.~\ref{sec:detector_concepts}.

\subsubsection{\tnu}
\label{sec:2v_bkg}
\begin{figure}[t]
    \centering
    \includegraphics[width=\columnwidth]{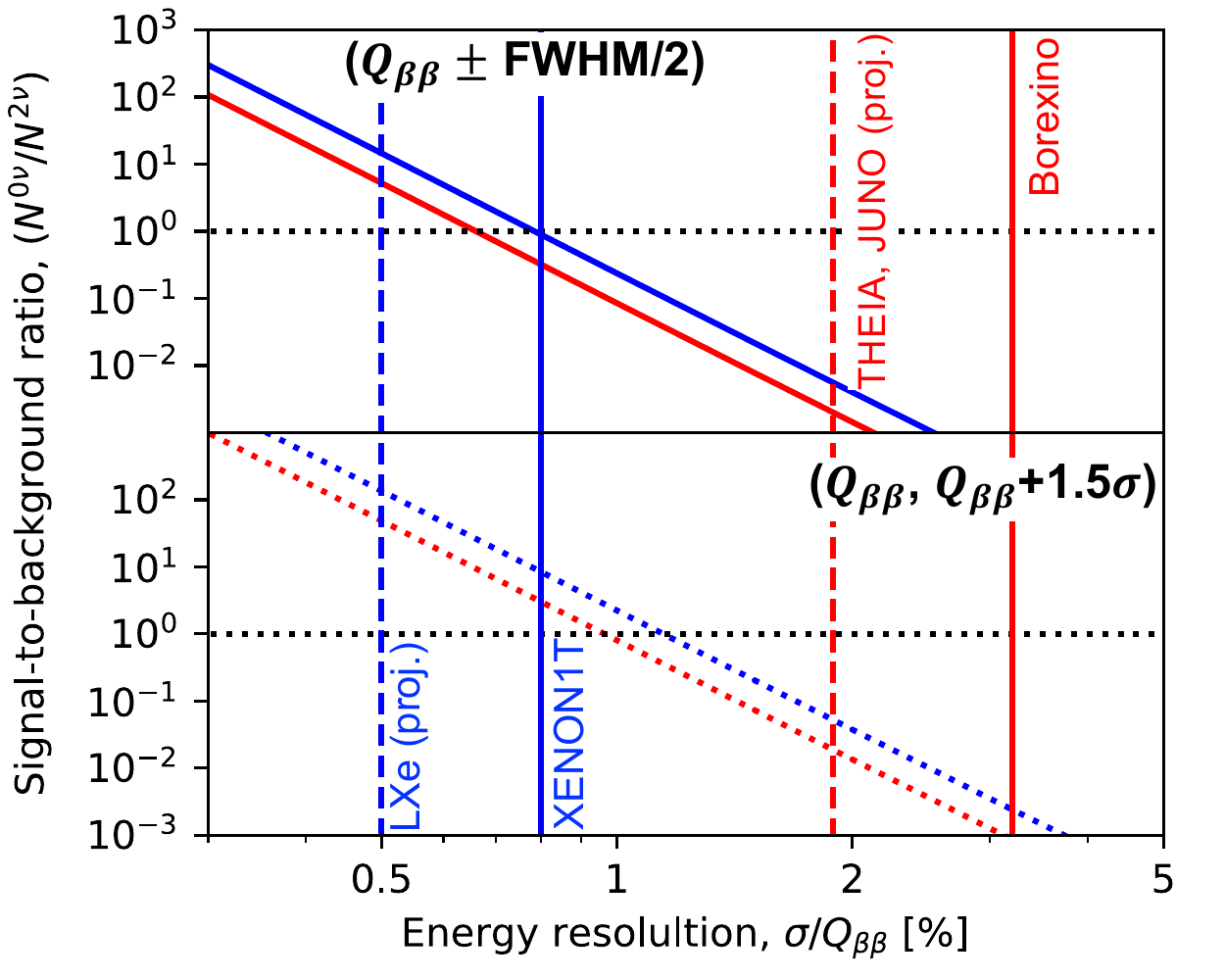}
    \caption{Signal-to-background ratio for \znu with half-life of $10^{30}$~yr, relative to backgrounds from the tail of the \tnu spectrum for a FWHM ROI centered on $Q_{\beta\beta}$ (top) and an asymmetric $(0,+1.5\sigma)$ region around $Q_{\beta\beta}$ (bottom)~\cite{Biller:2013wua_normal}. The ratio for \isoXe (blue) and $^{130}$Te are shown (red), which differ due to the factor of $\sim$3 difference in \tnu half-life~\cite{Barabash:2020nck_2v_t12}. The best demonstrated energy resolution for existing detectors (solid) and projected sensitivity (dashed) are also shown.}
    \label{fig:2v_bkgs}
\end{figure}

Backgrounds from the high-energy tail of the \tnu spectrum are reducible only through the energy resolution of the detector, since the signature for \tnu and \znu is otherwise identical. This remains true even for advanced strategies to remove backgrounds, e.g. by identifying the $^{136}$Ba daughter of the decay~(i.e., ``Ba-tagging'', e.g., \cite{Mong:2014iya_ba_tag,McDonald:2017izm_ba_tag,nEXO:2018nxx_ba_tag,Thapa:2020gjw_ba_tag,Rivilla:2020cvm_ba_tag}). Figure~\ref{fig:2v_bkgs} shows the expected signal-to-background ratio as a function of energy resolution for \isoXe and $^{130}$Te, which are candidate isotopes for large homogeneous detectors in which the \tnu background may be significant. Since the background due to \tnu scales approximately as $b_{2\nu} \propto \sigma^6$, where $\sigma$ is the detector energy resolution~\cite{Elliott_0v_review:2002xe}, even small improvements in the resolution can dramatically reduce the background.

The energy resolution in GXe detectors has been measured to be as low as $\sigma/E = 0.2$\% at $E=662$~keV and pressures up to 50~bar~\cite{1997NIMPA.396..360B_gxe_res}. When extrapolated to $Q_{\beta\beta}$, even better resolution is possible~\cite{Gomez-Cadenas:2019ges_gxe_review}. Achieving a resolution of $\sigma/Q_{\beta\beta} = 0.2$\% in a large GXe TPC would be more than sufficient to avoid \tnu backgrounds, and would appear off the left side of the plot in Fig.~\ref{fig:2v_bkgs}. 

Large LXe TPCs developed to date have poorer energy resolution than GXe TPCs, due to fluctuations in the fraction of the total energy in the ionization and scintillation channels and imperfect collection of the scintillation light~\cite{EXO-200:2003bso_anticorr,Szydagis:2011tk_NEST,EXO-200:2019bbx_exo200yields}. Nonetheless, the energy resolution already demonstrated in existing LXe TPCs such as XENON1T ($\sigma/Q_{\beta\beta} = 0.8$\%)~\cite{XENON:2020iwh_xen1T_resol} is sufficient to avoid \tnu backgrounds when considering an energy range $(0,+1.5\sigma)$ around $Q_{\beta\beta}$, rather than a FWHM region centered on $Q_{\beta\beta}$. 
As described in Sec.~\ref{sec:liquid_concept}, for a large LXe TPC optimized for resolution at $Q_{\beta\beta}$ and with negligible electronics readout noise, $\sigma/Q_{\beta\beta} = 0.5$\% should be achievable with light collection efficiencies $\gtrsim$10\%~\cite{EXO-200:2019bbx_exo200yields,XENON:2020iwh_xen1T_resol,nEXO:2021ujk_nexo_new_sens}. At this resolution, the \tnu background would also be sub-dominant over the FWHM region centered on $Q_{\beta\beta}$.

\subsubsection{Internal radiogenic backgrounds}
\label{sec:internal_bkgs}
In addition to the \tnu decay itself, any other radiogenic backgrounds that will not be attenuated by self-shielding must be removed from the Xe. For example, backgrounds from $^{222}$Rn are a significant contributor to the total background in ton-scale LXe TPCs for \znu and dark matter searches~\cite{nEXO:2021ujk_nexo_new_sens,DARWIN:2020jme_darwin0vbb}. Of particular concern is the decay of $^{222}$Rn daughters to $^{214}$Bi, which can decay with a branching ratio of 1.5\% via a $\gamma$ with energy of 2448~keV, within 0.4\% of $Q_{\beta\beta}$. Since Rn is a noble gas, it is more difficult to remove from the Xe using standard purification techniques and can continuously outgas from surfaces in the detector, plumbing, or purifier systems~\cite{Albert:2015nta_exo200radio,LZ:2020fty_LZ_radio,XENON:2020fbs_xenon_Rn}. 

Decays of Rn daughters in the Xe itself can be rejected with effectively 100\% efficiency by identifying coincident energy deposits. First, a $\beta$ is also emitted along with the 2448~keV $\gamma$, which will push the vast majority of such decays within the active detector region out of the energy region-of-interest. Any remaining decays (e.g. for which the $\beta$ falls below threshold) can be rejected by tagging the following $^{214}$Po $\alpha$ decay from its much higher light-to-charge ratio~\cite{EXO-200:2015ura_alpha_beta}. Assuming no improvement is made in the specific activity of $^{222}$Rn over ton-scale LXe detectors (where the measured or projected activity is $\sim 1$~$\mu$Bq/kg)~\cite{XENON:2020kmp_xenonnt,Mount:2017qzi_lz_tdr,nEXO:2021ujk_nexo_new_sens,xenonRnTAUP}, a rejection factor $\gtrsim 10^5$ is required to eliminate $^{214}$Bi decays through the $^{214}$Po coincidence in a kton-scale detector. Conservatively considering only the coincident $^{214}$Po $\alpha$, this requires a 100\% efficient veto for 2.8~ms following a candidate event (corresponding to $\sim$17$\times$ the half-life of 164~$\mu$s for the $^{214}$Po decay), which is straightforward to implement with negligible livetime loss~\footnote{We note that in ton scale detectors typically a $\sim$1~ms coincidence window is assumed~\cite{nEXO:2021ujk_nexo_new_sens,DARWIN:2020jme_darwin0vbb} leading to a rejection factor only $\gtrsim 10^3$, although this is straightforward to extend to several ms}.

Given the above rejection, the only significant radon-induced backgrounds then arise from $^{214}$Bi decays where, e.g., the $^{214}$Bi is plated on a surface such as the field rings or cathode, and the coincident $\alpha$ and $\beta$ deposit their energy only in inactive materials. In kton-scale TPCs, the effects of such $^{222}$Rn induced backgrounds are expected to be significantly mitigated relative to ton-scale experiments, since $\gamma$s originating from all such surfaces are attenuated by the same self-shielding factor shown in Fig.~\ref{fig:external_bkgs}. Naively extrapolating the same specific activity above of $\sim 1$~$\mu$Bq/kg to a kton-scale detector,
then the Rn-daughter plateout on the detector surfaces can produce external backgrounds comparable to the intrinsic detector material radiopurity of the Xe vessel itself (see Sec.~\ref{sec:detector_concepts}). In liquid detectors the self-shielding described in Sec.~\ref{sec:external_bkgs} is also sufficient to make this background negligible. In GXe detectors (at both the ton-scale and kton-scale), additional tagging of the coincident $^{214}$Bi $\beta$ from decays occurring on the cathode is estimated to be sufficient to make the Rn-induced external background sub-dominant to external materials backgrounds~\cite{NEXT:2020amj_NEXTtonne}.

Other radioimpurities that are not noble gases are expected to be efficiently removed by the {\em in situ} purification of the Xe, and have been found to be sub-dominant to $^{222}$Rn induced backgrounds in existing detectors. Nonetheless, more detailed studies are required to ensure no previously unobserved radioisotopes in existing detectors become dominant sources of background at the kton-scale. Here we assume that all such impurities can be sufficiently purified from the Xe source material prior to filling the detector, either through gas-phase heated getters for non-noble gas impurities, and distillation or gas chromatography for noble gas impurities, including $^{42}$Ar. 

\subsubsection{Cosmogenic backgrounds}
\label{sec:cosmo_bkgs}
We assume internal backgrounds (including those of cosmogenic origin) can be sufficiently purified from the initial Xe feedstock and focus here only on long-lived species that can be possibly created {\em in situ} during detector operation. The most prominent such cosmogenic background is $^{137}$Xe ($T_{1/2}$ = 3.8~mins), which produces $\beta$ decays with a $Q$-value of 4.2~MeV, providing a background at energies relevant for \znu. For GXe, topological rejection enables the single $\beta$ from $^{137}$Xe to be distinguished from the $\beta\beta$ signal (see Sec.~\ref{sec:gas_concept}). In addition, the production of $^{137}$Xe through capture of thermal neutrons in $^{136}$Xe($n,\gamma$)$^{137}$Xe reactions can also be identified from the coincident de-excitation $\gamma$s with a total energy of 4.03~MeV~\cite{Prussin:1977zz,EXO-200:2015edf_exo200_cosmo}. By tagging these de-excitation $\gamma$s, planned ton-scale LXe detectors are projected to mitigate backgrounds arising from $^{137}$Xe production within the TPC volume to $\lesssim 5$~evts/(FWHM kt yr)~\cite{nEXO:2021ujk_nexo_new_sens}. The livetime loss associated with this veto can be reduced by only vetoing a small spatial region of the detector. For example, in an LXe detector the neutral $^{137}$Xe is expected to move by only $\sim 2$~cm/(3.8 min), assuming recirculation with similar turnover time and temperature uniformity as existing detectors~\cite{EXO-200:2015ura_alpha_beta}. This movement allows a detector volume containing the expected drift even for several half-lives to be vetoed, while still constituting a small fraction of the total detector mass (and thus a negligible exposure loss).

Relative to ton-scale experiments employing enriched Xe, a reduction in the $^{137}$Xe background by $\gtrsim 10\times$ is sufficient to make this background sub-dominant in kton-scale detectors (assuming comparable depths, e.g. at SNOLAB~\cite{nEXO:2021ujk_nexo_new_sens}). Due to either the single $\beta$ rejection possible in GXe, or the improved containment of the de-excitation $\gamma$s in LXe (analogous to the improved self-shielding from external $\gamma$s described in Sec.~\ref{sec:external_bkgs}), this goal should be achievable. However, if required, $^{136}$Xe($n,\gamma$)$^{137}$Xe production can also be highly suppressed through the admixture of $\sim$10\% by volume of $^{131}$Xe (or, possibly, other noble elements with high neutron capture cross sections~\cite{NEXT:2020qup_he3}). Since the thermal $n$ capture cross section is roughly 2 orders-of-magnitude higher for $^{131}$Xe relative to \isoXe, the resulting number of captures on \isoXe can be correspondingly decreased. 
For an enriched detector, light isotopes such as $^{131}$Xe would be depleted from the \enrXe during enrichment, but could be separated from the enrichment tails and added back at $\lesssim$10\% concentration to sufficiently suppress any backgrounds. 

In principle other rare cosmogenic activation products not identified to date in large Xe detectors could be produced, e.g., by spallation of Xe or other detector materials~\cite{EXO-200:2015edf_exo200_cosmo, Cebrian:2020bwn, SuperCDMS:2018tqu, Saldanha:2020ubf}. Future work would be required to survey possible activation products of interest, in order to minimize risk that any such backgrounds may become significant at the kton-scale. However, the homogeneous nature of a large Xe detector generally allows such backgrounds to be discriminated from a \znu signal unless they produce only a single $e^-$ near $Q_{\beta\beta}$ (and no other correlated decay signatures). In GXe, the topological discrimination between $\beta$ and $\beta\beta$ events would provide further ability to identify and reject such possible backgrounds.

\subsubsection{Solar $\nu$ backgrounds}
\label{sec:solar_nu_bkgs}
While not a significant background for ton-scale detectors~\cite{nEXO:2021ujk_nexo_new_sens,NEXT:2020amj_NEXTtonne}, solar $\nu$ backgrounds become a substantial challenge at half-life sensitivities approaching $10^{30}$~yr. 
Charged current interactions produce highly multi-site signatures, and simulations of ton-scale detectors indicate that the fraction of charged current interaction events entering the single-site region-of-interest is $<10^{-7}$~\cite{nEXO:2017nam_nexo_2017_sens,nEXO:2021ujk_nexo_new_sens}, indicating that they are negligible even at the kton-scale.

In contrast, electron-neutrino elastic scattering (ES), $\nu + e^- \rightarrow \nu + e^-$, will produce a single $\beta$ that can mimic the localized energy deposits from $\beta\beta$ decays. Near $Q_{\beta\beta}$, the dominant source of such events arises from $^8$B solar $\nu$~\cite{deBarros:2011qq_solar_nu,Elliott:2017bui_solar_nu}. The rate of such events for a terrestrial detector is $\sim 0.2$~evts/(kt yr keV), roughly independent of the detector material~\cite{deBarros:2011qq_solar_nu}. This translates to a rate of $\sim 4.9\ (2.0)$~evts/[kt yr FWHM] at a relative resolution of $\sigma/Q_{\beta\beta} = 0.5$\% (0.2\%). This background is also uniformly distributed within the Xe, and is separable on an event-by-event basis from \znu decays in the same energy range only if single and double $\beta$s can be distinguished. 

Given the signal rate from Eq.~\ref{eq:0v_rate}, solar $\nu$ ES backgrounds require either: 1) enrichment of the Xe to enhance the ratio of $^{136}$Xe nuclei to electron scattering targets within the detector; 2) separation between $\beta$ and $\beta\beta$ decays near $Q_{\beta\beta}$; or 3) directional sensitivity to statistically separate solar $\nu$ ES originating from the direction of the sun from the isotropic angular distribution of $\beta\beta$ emission. The tradeoffs between these options, the cost of enrichment, and other considerations play a major role in the optimal detector concept, including gas or liquid phase operation and enrichment level, as described below. For example, GXe TPCs have already demonstrated the required single-$\beta$ rejection ($\gtrsim 10\times$) for a \natXe target through reconstruction of the $e^-$ topology~\cite{NEXT:2020amj_NEXTtonne}. Additionally, it may be possible to reconstruct the initial direction of the $\beta$ recoil in GXe, allowing further statistical discrimination. In either LXe or GXe detectors, some discrimination between $\beta$ and $\beta\beta$ decays may be possible from discriminators based on Cherenkov light~\cite{Brodsky:2018abk_cherenkov}. Finally, Ba-tagging with sufficiently high efficiency and selectivity could also be used to reject this background.

\subsection{Detector concepts}
\label{sec:detector_concepts}

\subsubsection{Liquid phase}
\label{sec:liquid_concept}
\begin{figure}[t]
    \centering
    \includegraphics[width=0.8\columnwidth]{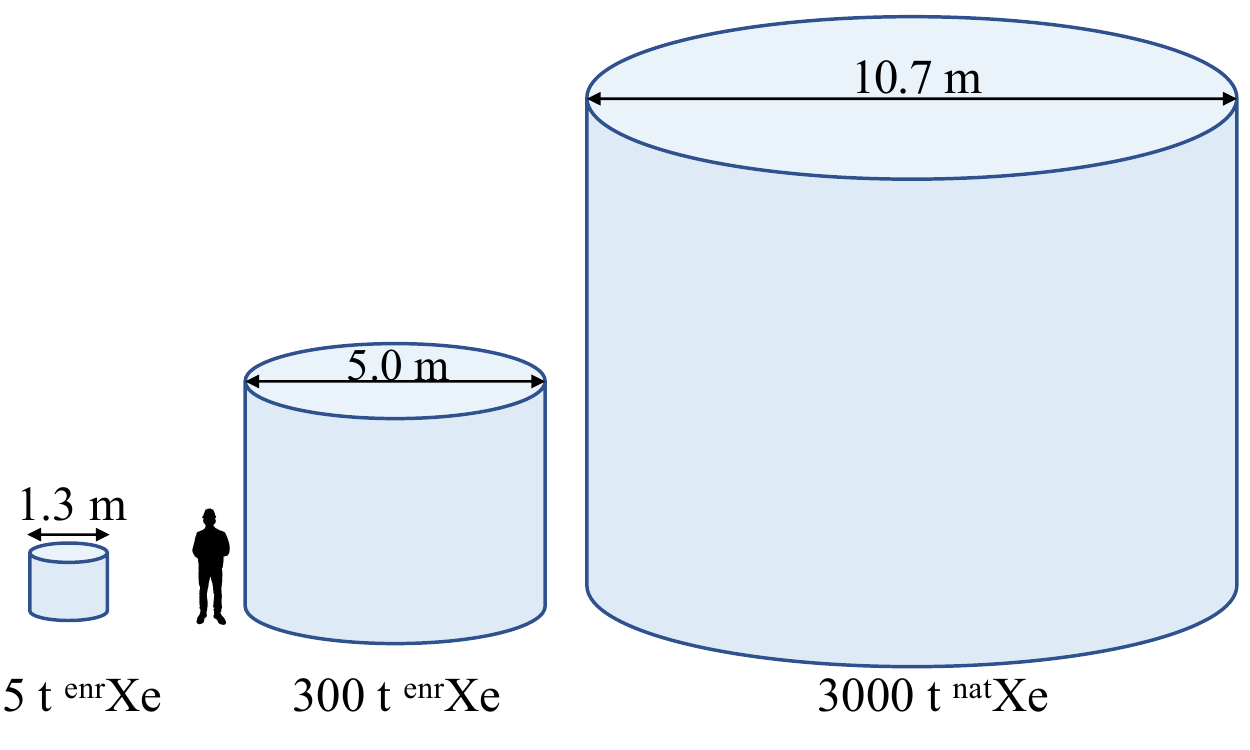}
    \caption{Schematic depiction of detector sizes for currently planned ton-scale liquid Xe TPCs for \znu (left, 5~t \enrXe)~\cite{nEXOpCDR:2018}, and the design concepts considered here that would be required to reach $\sim 10^{30}$~yr half-life sensitivity with either an enriched (center, 300~t \enrXe) or natural (right, 3000~t \natXe) liquid TPC.}
    \label{fig:lxe_det_sizes}
\end{figure}

A liquid phase detector would take advantage of the substantial self-shielding possible in a kton-scale detector. Optimal reduction of external backgrounds also dictates the ideal arrangement for the Xe, i.e., a single, homogeneous drift volume with nearly equal linear dimensions in all directions. Here we consider a cylinder with height equal to its diameter. As described below, sensitivity estimates have been performed to determine the size of such a detector that would be needed to reach the $10^{30}$~yr half-life sensitivity benchmark. In the following section we consider the two possible concepts shown in Fig.~\ref{fig:lxe_det_sizes}: an \enrXe detector (assuming 90\% enrichment fraction) with mass of 0.3~kt, and a \natXe detector with mass 3~kt, both of which contain approximately the same mass of \isoXe. 

\noindent{\em Energy resolution:} Existing LXe detectors have demonstrated energy resolutions as good as $\sigma/Q_{\beta\beta} = 0.8$\%~\cite{XENON:2020iwh_xen1T_resol}. As described in Sec.~\ref{sec:2v_bkg}, while this is sufficient to suppress leakage from \tnu backgrounds in the upper portion of the energy ROI, resolution models~\cite{EXO-200:2019bbx_exo200yields,nEXO:2021ujk_nexo_new_sens} indicates that the $\sigma/Q_{\beta\beta} = 0.5$\% target can be reached for a total light detection efficiency (i.e., the fraction of VUV scintillation photons producing a detected photoelectron (PE) in the light detector) of $>10$\%. Reaching this resolution in a large detector is accordingly driven by this light detection efficiency, provided other sources of noise such as readout noise in the charge and light channels remains sub-dominant~\cite{nEXO:2021ujk_nexo_new_sens,XENON:2020iwh_xen1T_resol}.

\noindent{\em Light collection:} Two concepts employed in existing detectors for light collection were studied for a kton-scale detector: 1) collection of light with photodetectors on only the flat faces of the cylinder, with a PTFE reflector around the barrel~\cite{Auger:2012gs_exo200jinst,Mount:2017qzi_lz_tdr,DARWIN:2016}; and 2) an optically open field cage with light detectors positioned around the TPC barrel~\cite{nEXO:2021ujk_nexo_new_sens,nEXO:2020mec_open_field_cage}. 
SiPMs can be used to directly detect Xe scintillation light with negligible readout noise~\cite{nEXO:2018cev_jamil_sipms,Gallina:2019fxt_gallina_sipms}, and in the coming years are likely to be combined with CMOS electronics into an integrated photon counter~\cite{Pratte:2021oqk_PDC}. A light propagation simulation of both designs 1 \& 2 above was performed in Chroma~\cite{Seibert2011FastOM_chroma} to determine the achievable light collection efficiency as a function of absorption length. Since the Rayleigh scattering length $\approx 30-50$~cm is much smaller than the linear dimensions of the detector, the light propagation is diffusive and photons transit a substantially larger linear distance than the detector size during propagation. Nonetheless, these simulations indicate that an absorption length of $\gtrsim 80$~m ($\gtrsim 40$~m) for designs 1 (2) is sufficient to reach the desired total $> 10$\% collection efficiency when combined with measured SiPM photon detection efficiencies~\cite{nEXO:2018cev_jamil_sipms,Gallina:2019fxt_gallina_sipms} and reflectivities~\cite{nEXO:2019jhg_sipm_refl,Lv:2019res_sipm_refl,nEXO:2021uxc_sipm_refl}. These absorption lengths are comparable to the lower limits extrapolated from existing measurements~\cite{Mount:2017qzi_lz_tdr,XENON:2015gkh_abs,BALDINI2005753_abs}, and are expected to improve with Xe purity. While light propagation over such long distances would need to be demonstrated, these estimates indicate that the required collection efficiencies should be feasible.

\noindent{\em Charge collection:} Charge collection in large liquid TPCs requires low readout noise ($\lesssim 600\ e^-$ per event) to ensure it is sub-dominant to the light collection in the overall resolution. This readout noise has been demonstrated in both existing single-phase or dual-phase designs~\cite{Auger:2012gs_exo200jinst,EXO-200:2019bbx_exo200yields,XENON:2020iwh_xen1T_resol}. In addition, a drift electric field $\gtrsim 200$~V/cm is expected to provide acceptable drift speeds for charge collection~\cite{EXO-200:2016qyl_diffusion}, while minimizing the effect of diffusion on the achievable topological signal/background discrimination. Achieving electron lifetimes $\gtrsim$20~ms (which has been recently demonstrated at the ton-scale with liquid phase purification~\cite{xenon_elifetime_TAUP}) would be sufficient to limit charge loss to $\lesssim$10\% in a kton-scale detector at the fields above. Diffusion effects are expected to be more significant at this scale than for ton-scale detectors, with an RMS smearing of 3.0~mm (4.2~mm) for charge drifting from the central region of a 300~t (3000~t) detector~\cite{EXO-200:2016qyl_diffusion,Njoya:2019ldm_diffusion,Hogenbirk:2018knr_diffusion}, which would be convolved with the initial 3--4~mm size of single cluster $\beta\beta$ decay events near $Q_{\beta\beta}$. 

For a 200~V/cm field, the required cathode voltage is $-100$~kV ($-215$~kV) for a single drift region in the 0.3~kt (3~kt) concepts shown in Fig.~\ref{fig:lxe_det_sizes}. These voltages are within a factor of $\sim$2 of the corresponding voltages in planned ton-scale detectors~\cite{nEXOpCDR:2018,Mount:2017qzi_lz_tdr}. While higher voltage operation of large LXe TPCs remains an area of active research~\cite{Rebel:2014uia_HV,Tvrznikova:2019xcg_HV}, these values are within plausible targets for HV possible in future detectors. Use of a central cathode (rather than single drift region) could also reduce the required voltages and effects from diffusion.

\noindent{\em Backgrounds:} The backgrounds described in Sec.~\ref{sec:backgrounds} were studied for the specific LXe detector concepts above. A Geant4~\cite{GEANT4:2002zbu} based simulation of backgrounds originating in the LXe vessel was performed to quantify the self-shielding of a large detector. This simulation assumes the dominant external $\gamma$ backgrounds arise from vessel (either due to internal or surface contamination), and uses the specific activity measured for commercially sourced copper (1~$\mu$Bq/kg for U/Th)~\cite{Auger:2012gs_exo200jinst,Leonard:2017okt_exo200assay,NEXT:2020amj_NEXTtonne}. The mass of the vessel was scaled from existing experiments by its surface area, assuming a thin-walled vessel supported by a fluid refrigerant as in existing ton-scale designs~\cite{nEXOpCDR:2018}. Backgrounds from the refrigerant are assumed to be sub-dominant to the vessel itself~\cite{nEXO:2021ujk_nexo_new_sens}. Surface backgrounds arising from daughters of $^{222}$Rn are similar in distribution to those in the vessel and are also included as external backgrounds.

Single-site versus multi-site separation was assumed to be comparable to existing ton-scale detectors, in which the rejection is sufficient to separate events within the $^{214}$Bi photoelectric interaction peak from Compton scatters with wider spacing (i.e. $\gtrsim$3~mm)~\cite{nEXO:2017nam_nexo_2017_sens}. The effect of diffusion on this rejection with longer drift distance remains to be studied in detail. However, even if the achievable SS/MS rejection is reduced relative to that assumed here, the required background level can still be reached by modestly increasing the standoff from the vessel walls (which in a large detector leads to only a small additional reduction in the fiducial mass). The results of this simulation indicate that a linear distance $>$42~cm from the vessel walls is sufficient to reduce the external $\gamma$ and $^{222}$Rn backgrounds to less than 10\% of the \znu decay rate from Eq.~\ref{eq:0v_rate}. As an example, for a 300~t detector approximately 57\% of the total mass (170~t) lies further than this distance from the vessel, while for a 3~kt detector this increases to 78\% of the detector mass (i.e., 2.3~kt of \natXe or 210~t of \isoXe).

$^{137}$Xe backgrounds are included after scaling the expected production rate per unit mass estimated for ton-scale detectors~\cite{nEXO:2021ujk_nexo_new_sens} by the improved vetoing that will be possible in a kton-scale LXe TPC. We assume a veto rejection inefficiency $\approx 10^{-3}$, which corresponds to the probability that one of the $\gtrsim$MeV de-excitation $\gamma$s from the production of $^{137}$Xe can exit through the 42~cm standoff from the vessel walls without interacting. A more detailed simulation of this vetoing would be expected to further improve the possible rejection efficiency, although this background is already sub-dominant for the conservative assumption above.

\begin{figure}[t]
    \centering
    \includegraphics[width=0.9\columnwidth]{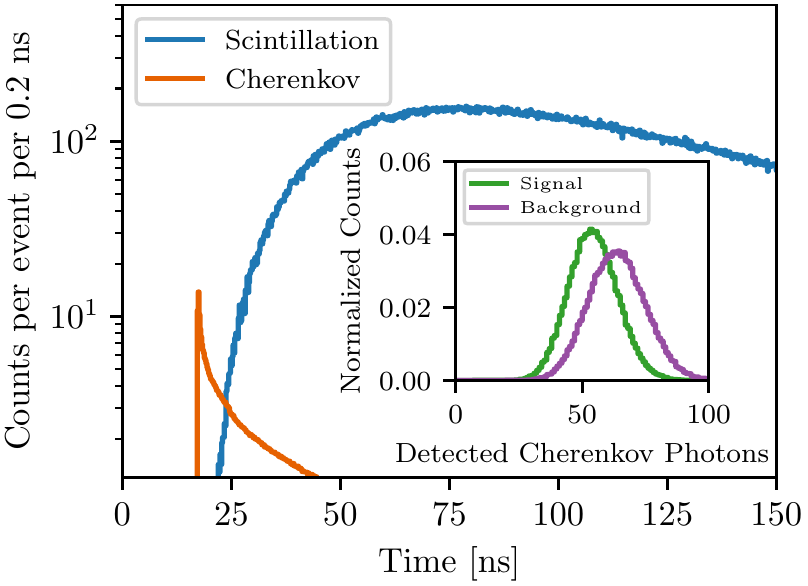}
    \caption{Simulated arrival time following the interaction of Cherenkov and scintillation photons for a kton-scale LXe detector. The inset shows the difference in expected detected Cherenkov photons for a single $e^-$ (background) and $\beta\beta$ (signal) at $Q_{\beta\beta}$. }
    \label{fig:cherenkov}
\end{figure}

For the volume of the detector that is greater than this standoff from the vessel walls, the dominant backgrounds arise from ES of $^8$B solar $\nu$ and the tail of the \tnu distribution, as described in Sec.~\ref{sec:backgrounds}. The $^8$B solar $\nu$ background is the primary challenge, especially in a \natXe target where the entire detector mass contributes to the backgrounds, while only a $\sim 10$\% mass fraction provides the signal. Reduction of this background may be possible through single-$\beta$ versus $\beta\beta$ separation based on the difference in the ratio of Cherenkov to scintillation light for the two event types~\cite{Brodsky:2018abk_cherenkov}. Cherenkov light can be separated from scintillation via timing. Simulations of a kton-scale detector indicate that the longer wavelength Cherenkov photons arrive primarily within $\lesssim$20~ns of the interaction time, prior to the arrival of the bulk of scintillation photons between 20~ns and several hundred ns (see Fig.~\ref{fig:cherenkov}). This timing resolution is easily within the capabilities of the integrated digital photon counters described above~\cite{Pratte:2021oqk_PDC}. The Chroma-based light simulation was also used to quantify the rejection that may be possible for the two light collection geometries considered. For the optimal timing-based $\beta$ vs. $\beta\beta$ separation of simulated 2.5~MeV events, a background acceptance of 35\% (i.e., a roughly $\sim3\times$ background rejection factor) was found at a \znu signal efficiency of 65\%. This rejection power was similar for both collection geometries and consistent with past simplified studies for kton-scale LXe TPCs~\cite{Brodsky:2018abk_cherenkov}. 

\begin{figure*}[t]
    \centering
    \includegraphics[width=\textwidth]{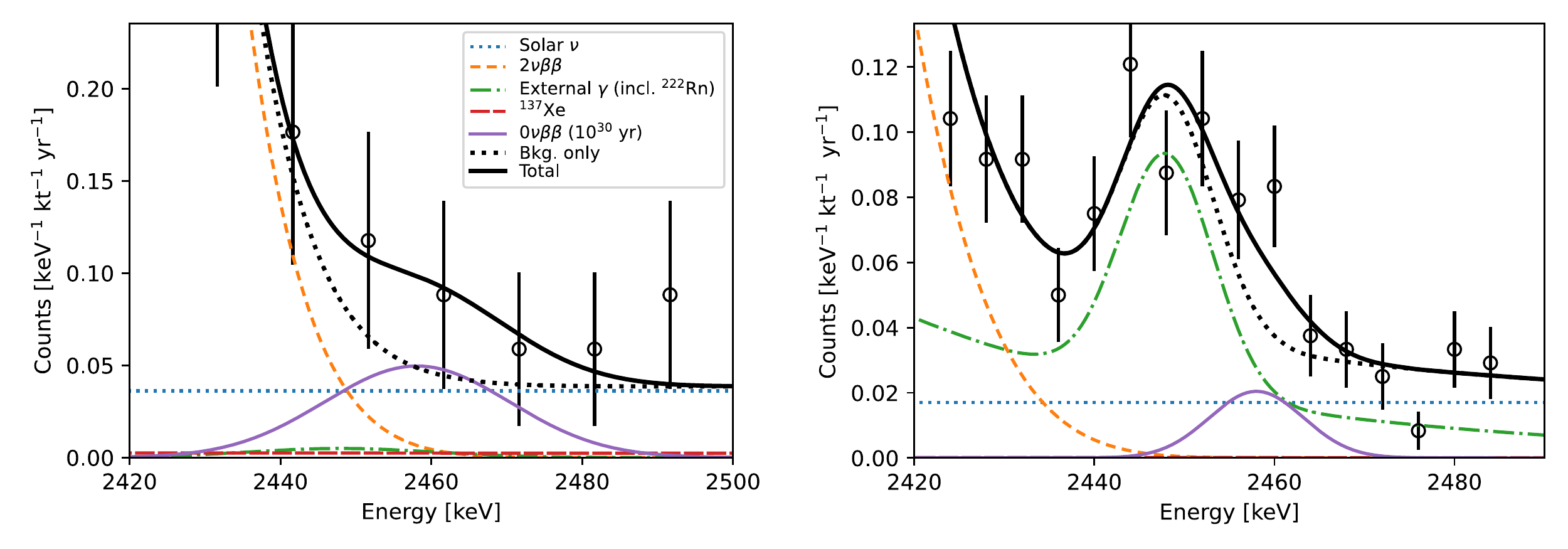}
    \caption{Example background model for a LXe concept (left) and GXe concept (right). For the LXe concept (left), the estimated spectrum is shown for an enriched detector in the fiducial region $>42$~cm from the vessel walls assuming $m_{\text{det}} = 300$~t, $\eta = 0.9$, $\sigma/Q_{\beta\beta} = 0.5$\% (fiducial mass 170~t), and Cherenkov-based single $\beta$ rejection with the efficiencies specified in the text. The GXe concept (right) assumes a \natXe detector with $m_{\text{det}} = 3$~kt, $\eta = 0.09$, $\sigma/Q_{\beta\beta} = 0.2$\%, with no fiducialization. In addition to the expected backgrounds, a potential \znu signal with a half-life of $10^{30}$~yr is shown. The error bars show an example toy dataset near the median discovery potential for a livetime of 20~yr. The units on the vertical axis are in terms of the fiducial detector mass (kt) and would require scaling by the enrichment factor to convert to isotope mass.}
    \label{fig:lxe_bkgs}
\end{figure*}

An example of the dominant estimated backgrounds in the central detector region are shown in Fig.~\ref{fig:lxe_bkgs} (left) for the 300~t \enrXe concept. The \natXe concept would have solar $\nu$ backgrounds that are roughly $10\times$ higher, but substantially reduced backgrounds from $^{137}$Xe and external $\gamma$s. For the assumed $\sigma/Q_{\beta\beta} = 0.5$\% resolution, the \tnu background is sub-dominant in the FWHM region around $Q_{\beta\beta}$.

\subsubsection{Gas phase}
\label{sec:gas_concept}
A GXe TPC at the kton scale was also considered. In comparison to the LXe concept, a GXe detector can more easily suppress the two irreducible backgrounds present at the kton-scale, i.e. ES of solar $\nu$ and the high energy tail of the \tnu distribution.
First, a GXe TPC can substantially suppress the solar~$\nu$ background by discriminating $\beta$ from $\beta\beta$ events through their topology. Tracks produced by single $e^-$ arising from a solar~$\nu$ ES in a gas TPC can be identified through a single high-density energy deposit (i.e., ``blob'') at the end of their track, while a \znu event would produce two blobs. Previous simulations have shown that with a gas pressure of 15~atm, this topological single $e^-$ discrimination could reject solar~$\nu$ backgrounds with 90\% efficiency~\cite{NEXT:2020amj_NEXTtonne}. In addition, GXe detectors at pressures $\lesssim 50$~bar avoid the event-by-event fluctuations in the deposited charge and light energy seen in LXe, enabling substantially better energy resolution and requiring only the deposited charge to be collected. This energy resolution is sufficient to fully eliminate the \tnu background if resolutions demonstrated in small scale detectors can be extended to the kton-scale. While no significant dependence of the energy resolution on density is expected for pressures between 15--50~bar~\cite{1997NIMPA.396..360B_gxe_res}, topological rejection is expected to degrade at pressures above the 15~bar pressure planned for ton-scale detectors~\cite{NEXT:2020amj_NEXTtonne}. Further work quantifying this topological rejection versus pressure is required to determine the optimal operating pressure for a kton-scale detector.

While the above backgrounds are substantially suppressed relative to the LXe design, the lower level of self-shielding due to the lower density in a GXe detector increases the impact of external backgrounds. External~$\gamma$s arising from the vessel materials become the dominant background in such a detector, and would be a primary driver of its design. 

\noindent{\em External backgrounds:} We consider a detector containing room temperature GXe at 15~atm, in the shape of a square cylinder to maximize self-shielding of the Xe. Optimizing the tradeoffs with higher pressure operation---which increases self-shielding and reduces the vessel size, but for which topological discrimination has not been studied in detail---are beyond the scope of the concepts considered here, but may provide more optimized designs. At this pressure, a Xe vessel radius of 12~m is required for a 1~kt detector (or 17~m for a 3~kt detector). A pressure vessel of such a diameter is likely to present a substantial engineering challenge and further study would be required to demonstrate its feasibility. However, solutions in which the cavern itself provides the mechanical support for a thin walled Xe vessel may be possible. In addition to conventionally mined caverns, such possibilities include use of a solution-mined salt cavern that would naturally support the required pressures~\cite{Monreal:2014gda_solution}.

To provide adequate shielding against external~$\gamma$s originating in the Xe pressure vessel, the vessel walls are assumed to be composed of three layers. Starting from the outside, a thick outer layer of stainless steel is assumed to maintain the high pressure internals (or, possibly, an alternative thinner vessel mechanically supported by the cavern walls). Regardless of the detailed design, backgrounds arising from the pressure vessel walls would be prohibitive if not shielded further. To shield external $\gamma$ radiation from the pressure vessel itself, a 2~m thick layer of ultra-pure and Rn-scrubbed water is assumed to surround the Xe. Geant4 simulations indicate this water thickness is sufficient to shield external $\gamma$s originating from the pressure vessel itself, such that the residual U/Th contamination in the water shield provides the dominant external background. We also assume that a thin nylon balloon~\cite{Benziger:2007iv} is placed between the water and the steel, to limit radon from the steel from emanating into the water. Finally a thin copper shell (with 2~mm thickness) is assumed to separate the water from the innermost region of GXe, with the same specific activity as assumed above for the LXe concept (1~$\mu$Bq/kg for U/Th). 

Due to the relatively small effect of self-shielding in the GXe design, alternative concepts employing multiple smaller modules with the same total mass might provide a more optimal design. In this scenario, improvement in material backgrounds by more than an order of magnitude relative to ton-scale detectors would be required to reach the required external background levels at half-lives $\sim10^{30}$~yr.

\noindent{\em Energy resolution:} Sufficient energy resolution $\sigma/Q_{\beta\beta} \lesssim 0.5$\% is required to avoid \tnu backgrounds. In addition, improved energy resolution can mitigate other broad spectrum backgrounds arising e.g., from solar $\nu$ and $^{137}$Xe. At relative resolutions $\lesssim 0.4$\%, separation between the \znu peak and the $^{214}$Bi $\gamma$ line at 2448~keV also starts to become possible, mitigating the dominant background from U contamination in external materials and $^{222}$Rn daughters on external surfaces or in inactive shielding. Although demonstrating that such resolution can be achieved in a kton-scale detector is still required, we assume here that $\sigma/Q_{\beta\beta} = 0.2$\% can be reached, which has already been demonstrated in small scale detectors even at energies substantially below $Q_{\beta\beta}$~\cite{1997NIMPA.396..360B_gxe_res}.

\begin{figure*}[t]
    \centering
    \includegraphics[width=\textwidth]{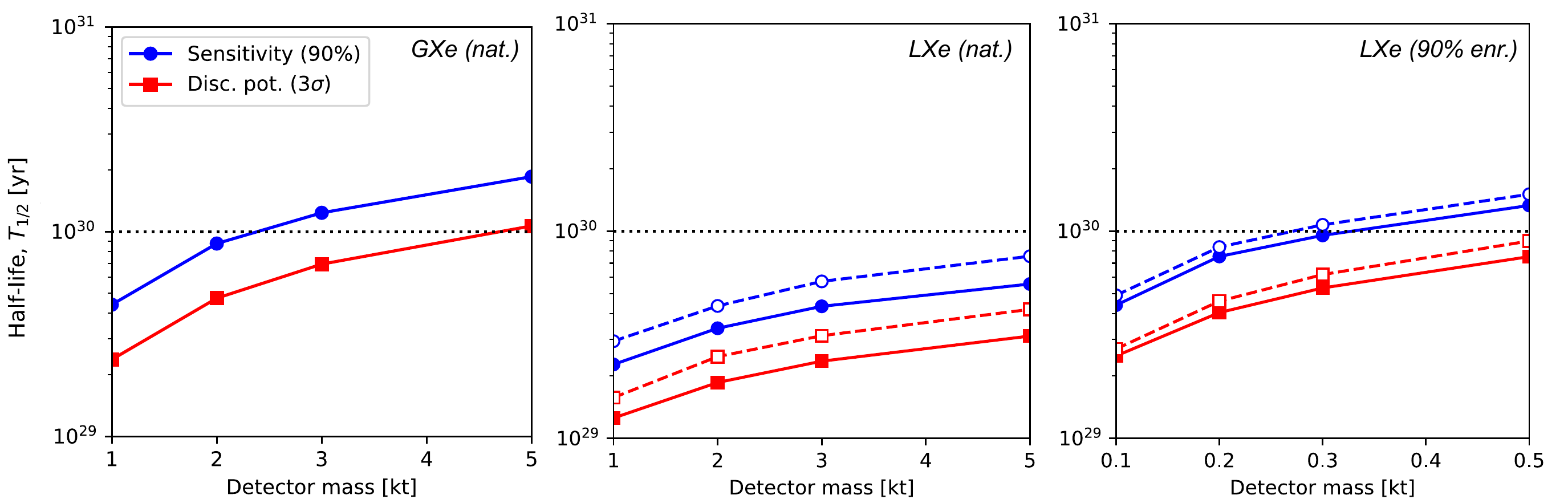}
    \caption{Estimated sensitivity versus detector mass for a \natXe GXe detector (left), a \natXe LXe detector (center), and an \enrXe LXe detector (right). The 90\% CL exclusion sensitivity (blue) and 3$\sigma$ discovery potential (red) are shown. For the LXe case where the solar $\nu$ backgrounds are more significant, the solid lines show the results with no Cherenkov-based single $\beta$ discrimination, while the dashed lines indicate the corresponding sensitivity using the Cherenkov based rejection efficiency described in the text. The benchmark half-life goal of $10^{30}$~yr sensitivity (dotted black) is reached for a $\sim$3~kt \natXe GXe detector and a $\sim$0.3~kt \enrXe LXe detector, while the \natXe LXe detector with the same sensitive mass reaches $\sim$40\% lower sensitivity due to the solar $\nu$ background.}
    \label{fig:lxe_sens_vs_mass}
\end{figure*}

\noindent{\em Charge collection:} Several possibilities exist for charge collection in a large GXe TPC. Existing GXe designs~\cite{NEXT:2015wlq_next100_sens, NEXT:2020amj_NEXTtonne} at the ton-scale employ charge amplification via electroluminescence (EL). Similar anode and cathode designs are in principle possible at the kton scale, although the required instrumented area becomes substantially larger than demonstrated to date. Maintaining the required topological rejection will likely require subdividing the detector volume into multiple drift regions to limit charge diffusion during drift. Such a design limits the required high voltage, at the cost of additional instrumented area and materials within the Xe volume. Detailed optimization of the number of drift regions, anode/cathode design, etc are beyond the scope of the concepts considered here, and we assume performance similar to ton-scale designs can be extended to the kton-scale.

A summary of the expected backgrounds for a 3~kton GXe detector employing \natXe following the concept above is shown in Fig.~\ref{fig:lxe_bkgs}~(right). Compared to the LXe case, external $\gamma$ backgrounds become more prominent due to the decreased self-shielding, while solar $\nu$ backgrounds are substantially reduced through the topological discrimination, avoiding the need for enrichment. The use of \natXe also suppresses the $^{137}$Xe background due to the natural presence of lighter isotopes such as $^{131}$Xe and $^{129}$Xe that capture the majority of thermal neutrons.

\subsection{Sensitivity} 

Based on the background models for the LXe and GXe concepts described in Sec.~\ref{sec:liquid_concept}--\ref{sec:gas_concept}, sensitivity studies were performed for both the \enrXe and \natXe concepts as a function of the detector mass and are shown in Fig.~\ref{fig:lxe_sens_vs_mass}. To calculate the exclusion sensitivity for each detector concept, toy Monte Carlo data sets were drawn from the background-only model and the 90\% CL lower limit on the half-life was determined from a fit to the toy datasets in the \znu region-of-interest (ROI) based on the profile of the negative log likelihood over the number of \znu counts. For simplicity the normalization of all background components in the fit were fixed and only the signal component was allowed to vary. This procedure provides a good approximation to a fit over the entire energy range, since sufficient statistics are available to determine the normalization of the background components with sub-dominant uncertainty from signal sidebands (in energy, topology, or distance from the detector walls)~\cite{nEXO:2021ujk_nexo_new_sens, NEXT:2020amj_NEXTtonne}. In addition to the exclusion sensitivity, the discovery potential was calculated following the same procedure to determine the half-life at which the no-signal hypothesis could be rejected at 3$\sigma$ by the median toy dataset, assuming a \znu signal were present. 

\begin{figure}[t]
    \centering
    \includegraphics[width=\columnwidth]{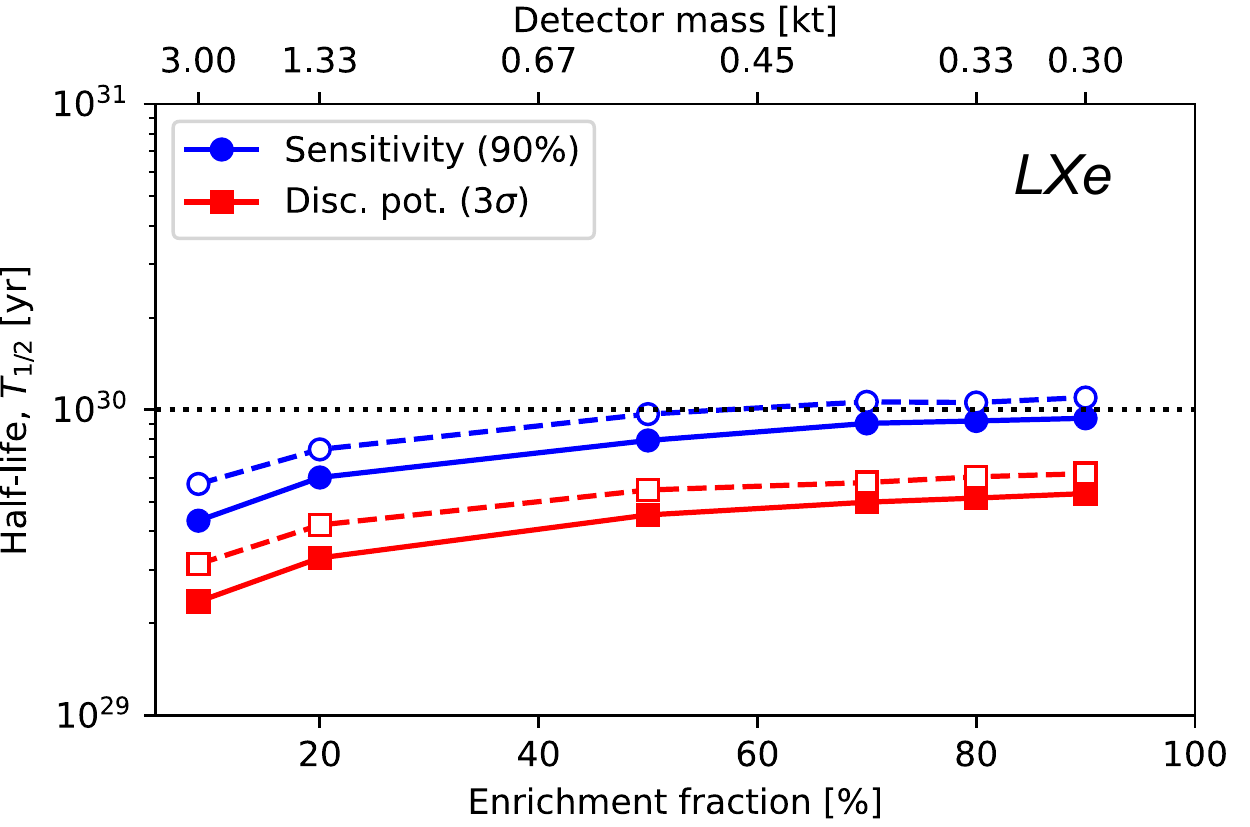}
    \caption{Sensitivity versus enrichment fraction for a LXe detector containing a fixed amount of \isoXe equal to the 300~t concept at 90\% enrichment fraction. The corresponding detector mass is indicated on the upper axis. Enrichment fractions above 50\% are sufficient to reach nearly optimal sensitivity. Dashed and dotted lines correspond to the same event selections as assumed in Fig.~\ref{fig:lxe_sens_vs_mass}. }
    \label{fig:lxe_sens_vs_enrich}
\end{figure}

Beyond the scaling with detector mass, the variation in sensitivity with various detector parameters was studied including enrichment fraction, energy resolution, and livetime. The GXe sensitivity was not found to vary strongly with enrichment since solar $\nu$ backgrounds were sub-dominant, although higher enrichment fractions permit a smaller overall detector size at the same sensitivity. In contrast, enrichment fractions $\gtrsim$50\% were found to be required for the LXe detector to reach an exclusion sensitivity $>10^{30}$~yr as shown in Fig.~\ref{fig:lxe_sens_vs_enrich}. For both concepts, the sensitivity follows a background limited scaling with livetime, $t$, at long times (i.e, $\sim \sqrt{t}$), with the bulk of the sensitivity achieved in $t=10$~yrs, but a 30\% relative increase in sensitivity for $t=20$~yrs operation. For the LXe (GXe) concepts, worsening the energy resolution relative to the baseline numbers assumed above still allowed a sensitivity $>0.8 \times 10^{30}$~yr to be achieved for a relative resolution $<0.7$\% ($<0.5$\%), respectively.

\subsection{Comparison to other technologies}
\label{sec:technol_comp}
The detector concepts and simplified sensitivity studies presented above indicate that either a kton-scale GXe or LXe detector may be able to reach sensitivities at, or near, the $10^{30}$~yr half-life benchmark. If Xe can be acquired in the required quantities, there are several advantages to incorporating it directly into a TPC relative to other possible detector technologies. In the context of the previous discussion in Secs.~\ref{sec:backgrounds}--\ref{sec:detector_concepts}, we briefly summarize those advantages here:
\begin{itemize}
    \item Modular detector designs based on Ge ionization detectors or cryogenic bolometers do not directly benefit from the self-shielding possible in homogeneous detectors since materials carrying backgrounds (detector supports, electronics, cabling, etc) are placed within the sensitive volume. Scaling to larger size thus does not directly reduce these backgrounds and substantial improvement in radiopurity of materials would be required relative to existing designs.

    \item Similar quantities of Xe could be doped into a large liquid scintillator detector (which could also employ $^{130}$Te, avoiding the isotope acquisition challenge for Xe). Such a detector would benefit from significant self-shielding and the ability to avoid external $\gamma$ backgrounds. However, the $\lesssim$0.5\% relative energy resolution needed to make the \tnu background negligible does not appear to be feasible in such a scheme. In addition, the typical loading fractions by mass of only 1--10\% result in relatively large solar $\nu$ backgrounds. The highest loading fractions possible with this method (even using \enrXe) are expected to be lower than achievable even for a \natXe TPC. Existing projections for such designs correspond to ultimate sensitivities between $10^{28}$--$10^{29}$~yr~\cite{Theia:2019,Cao:2019hli_juno,Zhao:2016brs_juno,Biller:2013wua_normal}.
    
    \item Ideas have been proposed to dope Xe into large LAr detectors at percent levels~\cite{snowmass_DUNE_beta}. While in principle possible, the increased  LAr mass relative to a Xe-only TPC would substantially increase the solar $\nu$ background. The presence of $^{42}$Ar is likely to also be a significant background in a large detector of this type~\cite{Barabash:2016lru_ar42,Lubashevskiy:2017lmf_ar42}. Finally, the larger detector size may not be optimal for reaching the required energy resolution.
    
    \item Alternative ideas using $^{82}$Se in an ion-drift TPC~\cite{NygrenSeF6:2018} or large array of pixellated sensors~\cite{ChavarriaSeCMOS:2016} may avoid the isotope acquisition challenges for Xe and might meet the resolution and background requirements. However, unlike large liquid noble TPCs these technologies are still under development and a detailed comparison with Xe TPCs is not yet possible.
\end{itemize}

\subsection{Alternative Xe-based concepts}
\label{sec:alternatives}
For simplicity, in this work we have focused on Xe acquisition and detector concepts capable of reaching the longest possible half-lives. However, intermediate scale detectors are possible and also can provide significant discovery potential. For example, an \enrXe detector with $\sim$50~t mass may be able to reach half-life sensitivities $\gtrsim 10^{29}$~yr. Production of the required Xe, either through the ideas presented here---or, at this scale, possibly from Xe captured from nuclear fuel reprocessing---may allow planned LXe detectors for dark matter~\cite{DARWIN:2016,DARWIN:2020jme_darwin0vbb} to be filled with \enrXe, probing portions of the allowed parameter space for \znu in the normal hierarchy. Other approaches include construction of a $\sim$300~t scale GXe or LXe TPC that could initially be filled with \natXe, running in parallel to the acquisition and enrichment of a similar quantity of \enrXe. Such an approach would provide a staged method for scaling to the ultimate sensitivity possible, while also lengthening the time over which Xe production can occur to minimize capital costs.

\section{Summary}
Acquisition of kton-scale quantities of Xe may enable rare-event searches with extreme sensitivity to \znu, dark matter, or other new BSM physics. Extensions to existing Xe TPC technology reaching sensitivity to \znu half-lives as long as $10^{30}$~yr appear plausible. The primary challenge to realizing such detectors is to acquire Xe in the required quantities. Since it appears infeasible to scale existing supply chains to the quantities needed for such a detector, fundamentally new methods for Xe acquisition may be required. In this work, we have described ideas for air capture of Xe using advanced adsorbent materials in a TSA process optimized for minimal energy consumption. While further R\&D is required to determine the feasibility of such an approach (or of other possible alternatives), studies to date suggest that capture of kton-scale quantities of Xe, potentially at reduced cost relative to existing methods, may be possible. If successful, an abundant and less-expensive supply of Xe would be likely to enable far reaching applications in both fundamental physics and beyond.

\begin{acknowledgments}
We would like to thank D.~Akerib, A.~Fan, B.~Jones, L.~Kaufman, and T.~Shutt for helpful discussions related to this work. This work was supported, in part, by the Department of Energy, Laboratory Directed Research and Development program at Lawrence Livermore National Laboratory, under contract DE-AC52-07NA27344.
\end{acknowledgments}

\bibliography{kton_xe}
\end{document}